\documentclass [prl,aps,letterpaper,twocolumn,superscriptaddress,amsmath,amssymb,floatfix,longbibliography] {revtex4-2}
\pdfoutput=1
\usepackage{graphicx}
\usepackage{textcomp}
\usepackage{mathptmx} 
\usepackage{amsmath}
\usepackage{xcolor}
\usepackage{soul}
\usepackage{comment}
\usepackage[hidelinks]{hyperref}
\usepackage{tabularx}
\usepackage{dcolumn}
\usepackage{ulem}
\usepackage{bm}
\usepackage{array}
\usepackage{indentfirst}
\usepackage{float}
\usepackage{subfigure}
\usepackage{algorithm}
\usepackage{algorithmicx}
\usepackage{algpseudocode}
\usepackage{verbatim}
\usepackage{makecell}
\usepackage{multirow}

\setcitestyle{super,sort&compress,open={},close={}}

\providecommand{\bibcommenthead}{}

\makeatletter
\AtBeginDocument{%
  \let\auto@bib\@empty
  \let\auto@bib@innerbib\@empty
}
\makeatother

\newcommand{\ket}[1]{\ensuremath{\left|#1\right\rangle}}

\definecolor{blue}{rgb}{0,0,1}
\definecolor{red}{rgb}{1,0,0}
\definecolor{green}{rgb}{0,1,0}

\newcommand{\HFRC}[1][Hefei National Research Center for Physical Sciences at the Microscale and School of Physical Sciences, University of Science and Technology of China, Hefei 230026, China]{\affiliation{#1}}
\newcommand{\SHRC}[1][Shanghai Research Center for Quantum Science and CAS Center for Excellence in Quantum Information and Quantum Physics, University of Science and Technology of China, Shanghai 201315, China]{\affiliation{#1}}
\newcommand{\HFNL}[1][Hefei National Laboratory, University of Science and Technology of China, Hefei 230088, China]{\affiliation{#1}}
\newcommand{\HNKL}[1][Henan Key Laboratory of Quantum Information and Cryptography, Zhengzhou, Henan 450000, China]{\affiliation{#1}}

\newcommand{\JIQT}[1][Jinan Institute of Quantum Technology and Hefei National Laboratory Jinan Branch, Jinan 250101, China]{\affiliation{#1}}

\newcommand{\QCTek}[1][QuantumCTek Co., Ltd., Hefei 230088, China]{\affiliation{#1}}
\newcommand{\SMX}[1][School of Microelectronics, Xidian University, Xi'an, China]{\affiliation{#1}}
\newcommand{\THU}[1][Department of Computer Science and Technology, Tsinghua University, Beijing, China]{\affiliation{#1}}
\newcommand{\PCL}[1][Pengcheng Laboratory, Shenzhen, Guangdong, China]{\affiliation{#1}}
\newcommand{\ZGL}[1][Zhongguancun Laboratory, Beijing, China]{\affiliation{#1}}

\begin{document}

\title{Surface code logical operations on a superconducting quantum processor}

\author{Weiping Lin}
\thanks{These authors contributed equally to this work.}
\HFRC
\SHRC
\HFNL
\author{Shaojun Guo}
\thanks{These authors contributed equally to this work.}
\HFRC
\SHRC
\HFNL
\author{Yuwei Ma}
\thanks{These authors contributed equally to this work.}
\HFRC
\SHRC
\HFNL
\author{Zhengzhong Yi}
\thanks{These authors contributed equally to this work.}
\HFRC
\SHRC
\HFNL
\author{Kai Zhang}
\THU
\PCL
\author{Jiahao Bei}
\SHRC
\author{Jianbin Cai}
\HFRC
\SHRC
\HFNL
\author{Sirui Cao}
\HFRC
\SHRC
\HFNL
\author{Danning Chen}
\QCTek
\author{Guoben Chen}
\SHRC
% ...

\begin{abstract}

Fault-tolerant quantum computation requires logical operations that manipulate encoded information while preserving quantum error-correction protection. In planar surface-code architectures, code deformation and lattice surgery provide a local, measurement-based route to such operations. Here we experimentally realize key elements of patch-based surface-code logical processing on a 107-qubit superconducting quantum processor. We first implement a reusable primitive layer comprising merge and split, patch expansion and shrinkage, and deformations mediated by domain walls and twist defects. We then compose these primitives to realize logical state routing, the logical controlled-NOT gate, and the single-qubit Hadamard and phase gates, which together form a Clifford-generating set. All operations are implemented on distance-three rotated surface-code patches with multi-round syndrome extraction and neural-network decoding, without post-selection. Our results advance superconducting surface-code experiments from protected logical memory to active, patch-based fault-tolerant logical operations.
\end{abstract}

\maketitle

Quantum computers are expected to solve problems beyond the 
reach of classical computation\cite{shorPolynomialTimeAlgorithms1997}, but useful large-scale algorithms 
require error rates far below those of directly controlled physical qubits\cite{gidneyFactor2048BitRSA2025}. 
Quantum error correction (QEC) addresses this challenge by encoding logical qubits 
nonlocally across many physical qubits\cite{shorSchemeReducingDecoherence1995}. The surface code is a leading 
architecture for solid-state quantum processors because of its high 
threshold and compatibility with two-dimensional local 
connectivity\cite{kitaevFaultTolerantQuantum2003,dennis2002topological,raussendorfFaultTolerantQuantum2007,raussendorfTopologicalFaultTolerance2007,fowler2012surface}. Recent superconducting-qubit 
experiments have demonstrated surface-code logical memories operating 
below threshold, where increasing the code distance suppresses the 
logical error rate\cite{google2025below,eickbuschDemonstrationDynamicSurface2025,he_experimental_2025}. 
These experiments establish protected storage as a critical milestone, 
but a fault-tolerant processor also requires logical operations that 
manipulate, entangle, and measure encoded states while maintaining 
error-correction protection throughout the operation.

In surface-code architectures, universal fault-tolerant quantum computation can be built from Clifford operations together with non-Clifford resources such as logical magic states\cite{bravyi2005universal}. The same locality constraints that make the surface code experimentally practical, however, restrict the logical gates available through transversal operations\cite{eastinRestrictionsTransversal2009,bravyiClassificationTopologically2013}. Code deformation and lattice surgery provide a measurement-based route around this limitation\cite{bombinQuantumMeasurements2009,horsman2012lattice}: in a patch-based architecture, logical operations are implemented by reconfiguring stabilizers, boundaries and joint parity measurements while the quantum information remains encoded. Logical qubits can then be moved, merged, split and deformed\cite{litinskiGameSurfaceCodes2019,fowler2019lowoverheadquantumcomputation}, and these primitives can be composed into a Clifford-generating set of logical gates\cite{brown2017poking,geherErrorcorrectedHadamardGate2023,gidney2024inplace}. Together with high-fidelity magic states\cite{litinskiGameSurfaceCodes2019,gidneyMagicStateCultivation2024,rosenfeldMagicStateCultivation2025}, they form a route to universal fault-tolerant computation.

Individual code-deformation and lattice-surgery operations have been demonstrated across ion-trap\cite{erhardEntanglingLogicalQubits2021a,ryan-andersonHighfidelityTeleportationLogical2024}, neutral-atom\cite{bluvsteinFaulttolerantNeutralatomArchitecture2026}, and superconducting platforms\cite{lacroixScalingLogicColour2025,besedinLatticeSurgeryRealized2026,wangSuperconductingSurfacecodeProcessor2026}. However, to our knowledge, no prior experiment has demonstrated both a unified code-deformation and lattice-surgery primitive layer and its composition into a Clifford-generating set of logical gates under repeated syndrome extraction and without post-selection.

Here we experimentally realize this patch-based surface-code logical-processing layer on a 107-qubit superconducting quantum processor. We first implement code-deformation and lattice-surgery primitives, then use them for logical state routing and compose them into logical CNOT, Hadamard and phase gates. All operations use distance-three rotated surface-code patches with multi-round syndrome extraction, neural-network decoding and no post-selection. We benchmark the routing and gate operations with decoded logical state and process tomography. Together, these results establish an active patch-based surface-code processing layer in a superconducting architecture.

\section*{Patch-based architecture and logical primitives}

\begin{figure}[!htbp]
	\includegraphics[width=\columnwidth]{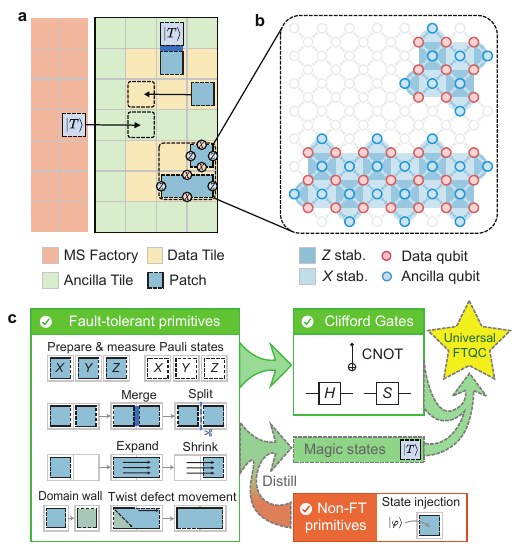}
	\caption{\textbf{Patch-based surface-code logical operations.}
	\textbf{a,} Tile-level layout of a patch-based surface-code processor. Peach, green and yellow tiles denote magic-state factories, ancilla tiles and data tiles, respectively; blue regions denote active logical patches used for routing and gate operations.
    \textbf{b,} Mapping of activated distance-three surface-code patches onto the superconducting-qubit lattice. Red and blue circles denote data and ancilla qubits, while dark-blue and light-blue plaquettes denote $Z$- and $X$-type stabilizers, respectively. Solid and dashed patch boundaries mark the two boundary types that define the logical Pauli operators.
	\textbf{c,} Schematic route to universal fault-tolerant quantum computation. Green and orange boxes distinguish fault-tolerant and non-fault-tolerant operations, respectively. Solid outlines and arrows mark the primitive and Clifford layers demonstrated here, while dashed elements indicate the magic-state resources, distillation steps and universal fault-tolerant quantum computation (FTQC) layer required for a full universal processor.}
	\label{fig1}
\end{figure}

At the layout level, surface-code computation is organized through a tile-and-patch abstraction (Fig.~\ref{fig1}a). The physical floorplan is coarse-grained into logical tiles that host surface-code patches, with each patch encoding one logical qubit and occupying either a single tile or a connected group of tiles. Each patch has two $X$-type and two $Z$-type boundaries, and strings of physical Pauli operators connecting the corresponding boundary pairs realize the logical operators $\bar{X}$ and $\bar{Z}$. Data patches store algorithmic logical qubits, ancilla patches mediate routing and joint parity measurements, and magic-state factories supply non-Clifford resource states\cite{litinskiGameSurfaceCodes2019,gidneyFactor2048BitRSA2025}.

This spatial abstraction becomes a computational architecture once patches can be prepared, deformed and measured through logical instructions (Fig.~\ref{fig1}c). In this work, we use lattice surgery to refer to merge--split parity measurements between patches, and code deformation to refer to changes in patch geometry or boundary structure that move or transform logical observables. The logical primitive layer consists of logical Pauli-state preparation and measurement, lattice-surgery merge and split operations, expansion and shrinkage, and deformations mediated by domain walls and twist defects. Merge and split operations temporarily join and separate patch boundaries to infer joint logical parities from newly activated stabilizer checks. Expansion enlarges a patch by activating additional physical qubits and stabilizer checks adjacent to a boundary, whereas shrinkage reduces the patch by measuring out a boundary region. Deformations mediated by domain walls and twist defects alter how Pauli strings are routed and transformed through the code. Sequences of these primitives form the logical Clifford layer. Lattice-surgery parity measurements provide entangling operations such as CNOT, while domain-wall and twist-defect deformations enable the single-qubit $H$ and $S$ operations. Clifford operations are not universal by themselves; universality is obtained by supplementing them with injected magic states that are promoted to high-fidelity logical resources by distillation or cultivation\cite{bravyi2005universal}. In this work, we focus on the primitive and Clifford layer shown with solid outlines in Fig.~\ref{fig1}c\cite{litinskiGameSurfaceCodes2019,bluntCompilationSimpleChemistry2024}.

The experiments were performed on a 107-qubit superconducting quantum processor\cite{he_experimental_2025,gao2025establishing,jiang2026one}. Device performance metrics are summarized in supplementary materials. Fig.~\ref{fig1}b shows how representative surface-code patch geometries are instantiated on the superconducting-qubit lattice. The upper patch is a standard distance-three rotated surface-code memory, while the lower patch is a rectangular rotated surface-code memory occupying two connected tiles, with asymmetric distances $d_X=3$ and $d_Z=7$. These calibrated patch geometries define the code units used in the logical operations below.

Before executing the compiled logical operations, we separately implemented and benchmarked the reusable merge, split, expansion and shrinkage protocols. These primitive-level benchmarks support the construction of the movement, CNOT, Hadamard and phase gates described below. The supplementary materials reports these benchmarks, together with detailed protocol schedules and circuit-distance verification for the compiled H, S and CNOT operations. For each compiled circuit reported in the main text, we sampled 100,000 shots, decoded the data with the neural-network decoder\cite{zhang2026learning}, and applied no post-selection. Parenthetical uncertainties for experimental quantities in the main text denote 95\% confidence-interval half-widths estimated by bootstrap resampling.

\section*{Logical movement}

We first test the primitive layer in a routing task, where an encoded state must be transferred between memory, interaction and magic-state-factory regions while retaining an encoded representation\cite{litinskiGameSurfaceCodes2019,gidneyFactor2048BitRSA2025}. We implemented and compared two strategies with closely related spacetime layouts: state teleportation by lattice surgery\cite{ryan-andersonHighfidelityTeleportationLogical2024,lacroixScalingLogicColour2025} and direct code deformation\cite{koboriLSQCAResourceEfficient2025}.

\begin{figure}[!htbp]
	\includegraphics[width=\columnwidth]{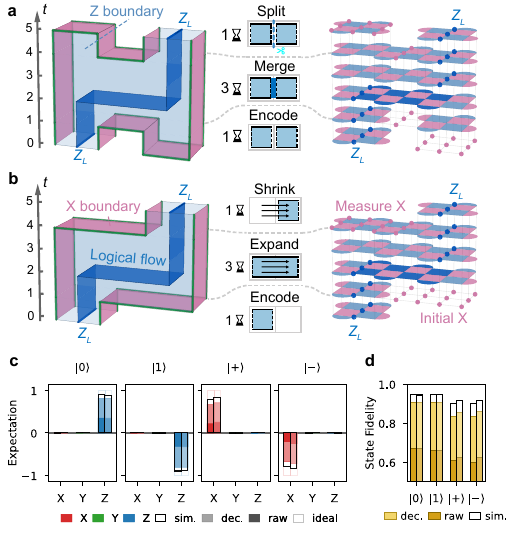}
    \caption{\textbf{Logical state routing by lattice surgery and code deformation.}
    \textbf{a,} Lattice-surgery teleportation. Left, spacetime diagram showing the transfer of $Z_L$ from the input patch to the auxiliary patch. Middle, sequence of lattice-surgery primitives. Right, implementation on the superconducting-qubit array, shown as stacked surface-code layers; dark-blue plaquettes mark newly established $Z$ stabilizers.
    \textbf{b,} Direct movement by code deformation. Left, spacetime diagram showing expansion of a single patch into the target region followed by shrinkage to the final patch. Middle, sequence of deformation primitives. Right, corresponding hardware layout and spacetime layers; removed data qubits are measured in the $X$ basis. In \textbf{a} and \textbf{b}, blue and pink faces denote $Z$ and $X$ boundaries, respectively.
    \textbf{c,}  Pauli expectation values for routed $\ket{0}$, $\ket{1}$, $\ket{+}$ and $\ket{-}$ states. Red, green and blue denote $X$, $Y$ and $Z$ observables; Filled bars show raw and decoded experimental values, and outlined bars show numerical simulations. For each input state, the left and right groups correspond to lattice-surgery teleportation and direct movement, respectively.
    \textbf{d,} Decoded logical-state fidelities for the two routing protocols.}
    \label{fig2}
\end{figure}

Fig.~\ref{fig2}a shows the state-teleportation protocol implemented by lattice surgery. The input patch was prepared in the target logical state, while an auxiliary patch was initialized in $\ket{+}$. A merge--split sequence measured the joint logical $ZZ$ parity between the two patches. The input patch was then measured in the $X$ basis, and the logical state was recovered on the auxiliary patch through Pauli-frame updates conditioned on the parity and $X$-basis measurement outcomes. 
Fig.~\ref{fig2}b shows the direct-movement protocol implemented by code deformation. A single surface-code patch was expanded into the target region, three QEC cycles on the enlarged rectangular code established the new stabilizers and extended the logical operators into the destination, and the original region was removed by measuring the excluded data qubits in the $X$ basis. The random outcomes of the newly introduced $Z$ stabilizers and removed-qubit measurements were incorporated into the Pauli frame.

Both protocols used the same $1+3+1$ QEC-cycle sequence and occupied a $3\times7$ spatial region in the experiment, highlighting their closely related spacetime structures. We benchmarked the two protocols by preparing four logical $X$- and $Z$-basis states, $\ket{0}$, $\ket{1}$, $\ket{+}$ and $\ket{-}$, and performing logical-state tomography after each routing operation (Fig.~\ref{fig2}c,d). For the lattice-surgery teleportation protocol, the decoded output state fidelities were $0.908(2)$ for $\ket{0}$, $0.911(4)$ for $\ket{1}$, $0.841(2)$ for $\ket{+}$ and $0.842(4)$ for $\ket{-}$. For the direct movement method, the corresponding decoded fidelities were $0.908(3)$, $0.905(1)$, $0.860(1)$ and $0.863(3)$.

Across both protocols, the $Z$-basis states retained higher fidelity than the $X$-basis states. The two protocols gave nearly identical fidelities for $\ket{0}$ and $\ket{1}$, whereas the direct-movement protocol improved the $\ket{+}$ and $\ket{-}$ fidelities by about $0.02$. A quantitative analysis of this basis dependence and protocol comparison is provided in the supplementary materials.

\section*{Logical CNOT gate}

\begin{figure*}[t]
	\includegraphics[width=\textwidth]{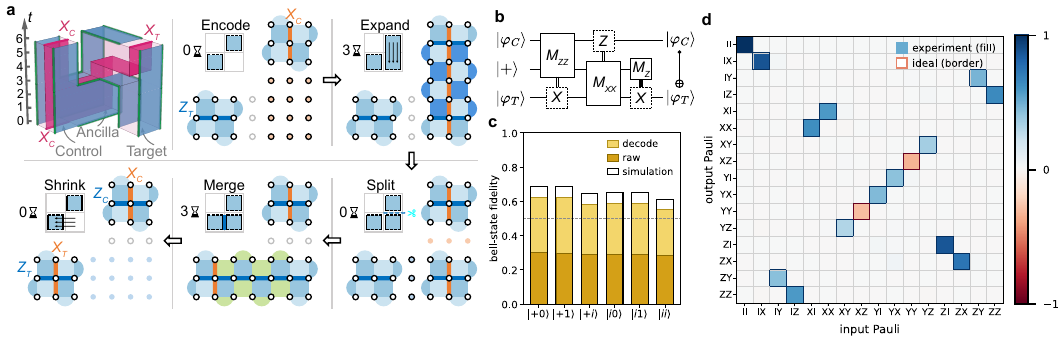}
    \caption{\textbf{Logical CNOT gate by lattice surgery and code deformation.}
    \textbf{a,} Spacetime structure and patch sequence. Left, spacetime diagram for the control, target and ancilla regions, showing the propagation of $X_C$ into the output operator $X_CX_T$. Right, physical patch layouts on the qubit grid during initialization, expansion, splitting, merging and shrinkage. The blue and orange strings track the transformations $Z_T\rightarrow Z_CZ_T$ and $X_C\rightarrow X_CX_T$, respectively. Dark-blue and green plaquettes indicate newly established $Z$- and $X$-type stabilizers.
    \textbf{b,} Equivalent logical circuit with joint measurements $M_{ZZ}$ and $M_{XX}$, ancilla readout $M_Z$, and Pauli-frame updates.
    \textbf{c,} Bell-state fidelities for six input product states. Filled bars show raw and decoded experimental values, and outlined bars show numerical simulations.
    \textbf{d,} Reconstructed two-logical-qubit Pauli transfer matrix (PTM). Filled squares show experimental PTM elements, red outlines mark the ideal CNOT entries and the colour scale gives the signed PTM amplitude.}
    \label{fig3}
\end{figure*}

We next assemble the same primitive layer into an entangling logical instruction. The CNOT gate transforms the logical Pauli operators as $X_C\rightarrow X_CX_T$, $Z_T\rightarrow Z_CZ_T$, while leaving $Z_C$ and $X_T$ unchanged. In planar surface-code layouts, this transformation is naturally implemented by lattice surgery through joint logical parity measurements\cite{horsman2012lattice}. We implemented the gate using the same expansion, split, merge and shrink primitives used above for logical state routing.

Fig.~\ref{fig3}a shows the operation as a spacetime protocol involving control, target and ancilla regions. The spacetime diagram tracks the nontrivial observable flow, $X_C\rightarrow X_CX_T$. In the distance-three implementation, the compiled schedule uses a $7\times7$ spatial region and six syndrome-extraction cycles. 
The corresponding patch schedule is shown in the right panel of Fig.~\ref{fig3}a. The control patch is first expanded into the ancilla region, followed by a split operation that separates the ancilla from the control. The target patch is then merged with the ancilla and finally shrunk back to the target region. In the circuit representation (Fig.~\ref{fig3}b), these primitives realize the required joint logical measurements, $M_{ZZ}$ and $M_{XX}$, followed by a logical $Z$-basis measurement of the ancilla. The CNOT operation is completed by Pauli-frame updates conditioned on these measurement outcomes.

We characterize the gate in two complementary ways. First, we prepare six input product states that ideally produce Bell states after the CNOT. The decoded Bell-state fidelities range from $0.550(1)$ to $0.624(2)$ (Fig.~\ref{fig3}c); all six values exceed the separability bound of $0.5$, witnessing entanglement between the output logical patches\cite{guhneEntanglementDetection2009}. Second, we perform two-logical-qubit process tomography\cite{chuangPrescriptionExperimentalDetermination1997}, with logical $Y$-basis access implemented by state injection\cite{yeLogicalMagicState2023}. The reconstructed Pauli transfer matrix (PTM) (Fig.~\ref{fig3}d) displays the ideal CNOT signed-permutation structure, with an average logical gate fidelity of $0.643(2)$.

Both the ideal-support PTM amplitudes and the input-state dependence of the Bell-state fidelities reveal a hierarchy set by observable-flow spacetime volume and logical $Y$-basis access used in tomography. The invariant observables that remain stationary on their output patches yield the largest PTM entries ($ZI\rightarrow ZI$ and $IX\rightarrow IX$). Observables whose flows pass through expansion, split, merge or shrinkage segments have smaller PTM amplitudes, and PTM entries involving logical $Y$ operators are further attenuated by the state-injection-based $Y$-basis state preparation and measurement (SPAM) operations used for tomography. The Bell-state fidelities show the same ordering. Calibrated Pauli simulations reproduce this trend, and simulations with tomography-boundary SPAM removed substantially reduce the spread between input states. A quantitative analysis, including benchmarks of the Pauli-basis SPAM protocols, is provided in the supplementary materials.

\section*{Logical H gate}

Single-qubit Clifford gates require deformations beyond ordinary lattice-surgery parity measurements. The logical Hadamard gate exchanges the Pauli axes, mapping $X_L\leftrightarrow Z_L$ and $Y_L\rightarrow -Y_L$. In a surface-code patch, a transversal layer of physical $H$ gates exchanges the Pauli basis, but also interchanges the $X$- and $Z$-type boundaries and leaves the patch in a rotated orientation\cite{horsman2012lattice,geherErrorcorrectedHadamardGate2023}. We therefore combine a geometric patch rotation with a final transversal-$H$ domain wall, so that the logical Pauli axes are exchanged while the output patch has the intended orientation.

\begin{figure}[!htbp]
    \includegraphics[width=\columnwidth]{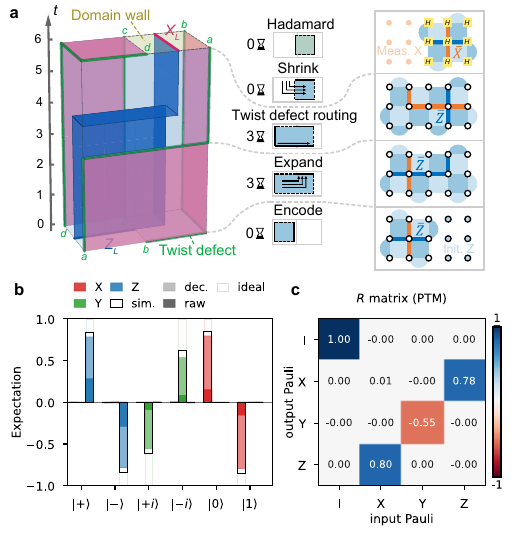}
    \caption{\textbf{Logical Hadamard gate by geometric patch rotation and a domain wall.}
    \textbf{a,} Spacetime structure and physical implementation. Left, spacetime diagram showing $Z_L\rightarrow X_L$ after the patch deformation and final Hadamard domain wall. The green lines trace the logical-corner trajectories, whose permutation $abcd\rightarrow dabc$ gives the $90^\circ$ geometric patch rotation. Middle, protocol sequence consisting of expansion, twist defect routing, shrinkage and a final transversal-$H$ layer. Right, implementation on the superconducting-qubit grid, showing the complementary observable flow $X_L\rightarrow Z_L$.
    \textbf{b,} Pauli expectation values for the six Pauli eigenstate inputs. Red, green and blue denote $X$, $Y$ and $Z$ observables. Filled bars show raw and decoded experimental values, and outlined bars show numerical simulations.
    \textbf{c,} Reconstructed logical PTM. The colour scale gives the signed PTM amplitude, and the dominant entries show the Hadamard mappings $X\rightarrow Z$, $Z\rightarrow X$ and $Y\rightarrow -Y$.}
    \label{fig4}
\end{figure}

Fig.~\ref{fig4}a shows the spacetime representation and implementation sequence of the logical Hadamard gate. In the spacetime diagram, a tracked $Z_L$ observable follows the deformed boundary frame and exits as $X_L$ after crossing the final Hadamard domain wall. The middle panel decomposes the implementation into surface-code primitives. The patch is first expanded for three syndrome-extraction rounds, followed by three rounds of logical-corner routing. The subsequent shrink step and the final transversal-$H$ layer add no additional syndrome-extraction round. The routed logical corners are twist-defect world lines where \(X\)- and \(Z\)-type boundary sheets meet; their permutation $abcd\rightarrow dabc$ represents a $90^\circ$ rotation of the patch boundary frame. For $d=3$, the compiled operation uses a $3\times5$ spatial footprint and six syndrome-extraction cycles.

Decoded state tomography verifies the expected Hadamard Pauli flow over the six Pauli eigenstate inputs (Fig.~\ref{fig4}b). The $X$-basis inputs $\ket{+}$ and $\ket{-}$ were mapped to $Z$-basis outputs with decoded expectations $\langle Z\rangle=0.785(7)$ and $-0.783(5)$, respectively. The $Z$-basis inputs $\ket{0}$ and $\ket{1}$ were mapped to $X$-basis outputs with $\langle X\rangle=0.797(7)$ and $-0.796(4)$. The $Y$-basis inputs showed the expected sign reversal, with $\langle Y\rangle=-0.557(4)$ and $0.536(9)$ for $\ket{+i}$ and $\ket{-i}$, respectively.
The reconstructed logical Pauli transfer matrix in Fig.~\ref{fig4}c shows the same dominant structure, with $R_{X\rightarrow Z}=0.796(2)$, $R_{Z\rightarrow X}=0.784(5)$ and $R_{Y\rightarrow Y}=-0.546(4)$. The corresponding average logical gate fidelity is $0.854(1)$.

The smaller $Y\rightarrow -Y$ amplitude has two physical origins. First, the $Y$-basis tomography settings use state-injection-based $Y$-basis SPAM, which attenuates the corresponding amplitude more strongly than transversal $X/Z$-basis SPAM. Second, an encoded $Y_L$ eigenstate is sensitive to logical failures in both the $X_L$ and $Z_L$ channels. Simulations with tomography-boundary SPAM removed isolate the SPAM contribution; the remaining attenuation is consistent with the sensitivity of a $Y_L$ eigenstate to both $X_L$- and $Z_L$-type logical failures. Quantitative details are provided in the supplementary materials.

\section*{Logical S gate}

We next implement the logical phase gate. Together with the Hadamard gate, $S_L$ generates the single-qubit Clifford group. The logical operation $S_L=\mathrm{diag}(1,i)$ maps
$X_L\rightarrow Y_L$, $Y_L\rightarrow -X_L$, and leaves $Z_L$ invariant. Efficient implementations of $S_L$ are particularly important because conditional logical $S$-gate corrections arise naturally in logical T gate teleportation protocols\cite{chamberlandCircuitLevelProtocol2022,bombin2023logical,gidney2024inplace,hiraiSpacetimeVolumeLogicalS2026}. Here we realize $S_L$ through logical gate teleportation using a fault-tolerantly prepared $\ket{+i}_L$ ancilla.

\begin{figure}[!htbp]
    \includegraphics[width=\columnwidth]{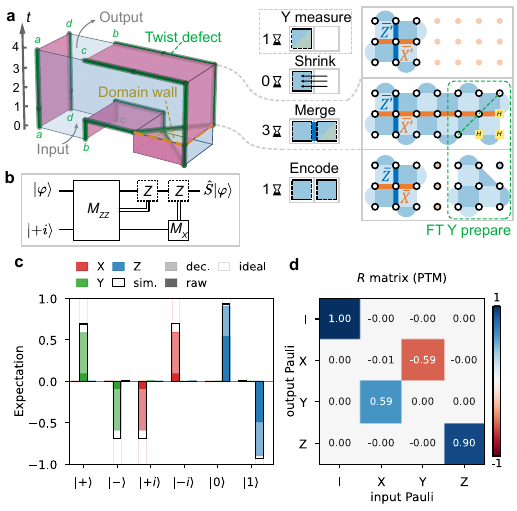}
    \caption{\textbf{Logical \(S\) gate by gate teleportation with a fault-tolerant \(Y\)-basis ancilla.}
    \textbf{a,} Spacetime structure and physical implementation. Left, in-place preparation of a logical \(\ket{+i}_L\) ancilla by diagonal twist defect motion, with corner permutation \(abcd\rightarrow acbd\). Middle, protocol sequence: parallel encoding, three-round lattice-surgery merge, shrinkage and optional fault-tolerant \(Y_L\)-basis readout. Right, implementation on the superconducting-qubit grid; the dashed green region marks the in-place \(Y_L\)-basis preparation.
    \textbf{b,} Gate-teleportation circuit. A joint logical \(M_{ZZ}\) measurement and ancilla \(M_X\) measurement teleport \(\ket{\varphi}\) to \(S\ket{\varphi}\), up to Pauli-frame \(Z\) updates.
    \textbf{c,} Pauli expectation values for six Pauli eigenstate inputs.  Red, green and blue denote $X$, $Y$ and $Z$ observables; filled bars show raw and decoded experimental values, and outlined bars show numerical simulations.
    \textbf{d,} Reconstructed logical PTM. The colour scale gives the signed PTM amplitude, and the dominant entries show the \(S\)-gate mappings \(X_L\rightarrow Y_L\), \(Y_L\rightarrow -X_L\) and \(Z_L\rightarrow Z_L\).}
    \label{fig5}
\end{figure}

The spacetime implementation is shown in Fig.~\ref{fig5}a. In the compiled distance-3 implementation, the key ingredient is in-place fault-tolerant access to the logical $Y$ basis, following the diagonal-twist construction of ref.~\citenum{gidney2024inplace}. Building on the Hadamard-domain-wall picture introduced above, the wall now terminates in the code bulk rather than spanning the whole patch. In the bulk, the boundary of the domain-wall surface corresponds to the world line of a twist defect. The resulting diagonal trajectory exchanges the two corners bordering the $X$-type boundary $bc$, giving the patch-corner permutation $abcd\rightarrow acbd$ and realizing the geometric action of the logical $S$ transformation\cite{brown2017poking,bombin2023logical}. Equivalently, this diagonal twist-defect motion prepares the logical $\ket{+i}_L$ ancilla within the original patch footprint while preserving the verified circuit distance.
The compiled protocol first encodes the input patch and the $\ket{+i}_L$ ancilla in parallel for one syndrome-extraction cycle, performs a three-round lattice-surgery merge, absorbs the shrink step without an additional syndrome-extraction round, and completes the teleportation readout with a logical $X$-basis measurement. For $d=3$, the compiled operation occupies a $3\times 7$ spatial footprint. Details of the circuit construction are provided in the supplementary materials.

In the circuit representation (Fig.~\ref{fig5}b), the input state $\ket{\varphi}$ is coupled to the fault-tolerantly prepared $\ket{+i}_L$ ancilla through a joint logical $M_{ZZ}$ measurement. A subsequent logical $M_X$ measurement of the ancilla completes the teleportation, and the two binary outcomes determine virtual $Z$-frame updates. After these Pauli-frame corrections, the remaining output patch carries $\hat{S}\ket{\varphi}$.

Decoded state tomography verifies the expected Pauli flow (Fig.~\ref{fig5}c). The $X$-basis inputs $\ket{+}$ and $\ket{-}$ were mapped to the $Y$ basis, with decoded expectations $\langle Y\rangle=0.584(7)$ and $-0.590(4)$, respectively. The $Y$-basis inputs $\ket{+i}$ and $\ket{-i}$ were mapped to the opposite $X$ basis, giving $\langle X\rangle=-0.590(6)$ and $0.588(3)$. The $Z$-basis inputs were preserved more strongly, with $\langle Z\rangle=0.903(5)$ for $\ket{0}$ and $-0.899(2)$ for $\ket{1}$. 
The reconstructed Pauli transfer matrix in Fig.~\ref{fig5}d shows the same dominant structure: $R_{X\rightarrow Y}=0.589(3)$, $R_{Y\rightarrow X}=-0.587(3)$ and $R_{Z\rightarrow Z}=0.901(3)$. The corresponding average logical gate fidelity is $0.846(1)$.

The $Z_L\rightarrow Z_L$ PTM transfer amplitude is significantly larger than the $X_L\rightarrow Y_L$ and $Y_L\rightarrow -X_L$ transfer amplitudes. This is because the $Z_L$ operator remains largely on the data patch and is protected by the long direction of the intermediate rectangular code and by transversal $Z$-basis SPAM. The $X_L$ and $Y_L$ operators, by contrast, are routed through the ancilla-mediated teleportation path, which incorporates fault-tolerant deformation-based $Y$-basis access. At the present code distance and physical error rates, this access remains more costly than transversal $Z$-basis SPAM. Unlike the $H$ and CNOT characterizations, where $Y$-basis access is used only for auxiliary tomography and its contribution can be separated in analysis, the fault-tolerant $\ket{+i}_L$ ancilla is an intrinsic part of the $S$ gate itself. A quantitative analysis is provided in the supplementary materials.
\section*{Discussion and Outlook}

We have demonstrated a patch-based surface-code logical-operation layer on a superconducting quantum processor. Using distance-three rotated surface-code patches with repeated syndrome extraction and decoding, we implemented reusable code-deformation and lattice-surgery primitives, including merge and split, expansion and shrinkage, and deformations involving domain walls and twist defects. We then composed these primitives into logical state routing and a Clifford-generating set comprising logical Hadamard, phase and controlled-NOT gates. All operations were performed without post-selection, and decoded state and process tomography verified the intended logical Pauli transformations.

The main advance is architectural as well as experimental. In a scalable patch-based computer, logical gates are spacetime protocols in which stabilizers, boundaries and logical Pauli representatives are reconfigured while the quantum information remains encoded\cite{bombin2023logical,litinskiGameSurfaceCodes2019,fowler2019lowoverheadquantumcomputation}. Our experiment shows that these protocols can be compiled onto a fixed two-dimensional superconducting layout, calibrated, decoded and benchmarked as modular logical instructions. At the same time, the present implementation is a finite-size distance-three demonstration. The reported process-tomography fidelities include physical-operation errors accumulated over the logical-circuit spacetime volume, together with tomographic SPAM contributions. Future SPAM-insensitive logical benchmarking could isolate the intrinsic logical-operation errors more directly\cite{lacroixScalingLogicColour2025,combesLogicalRandomizedBenchmarking2017}.

The next experimental milestone is to demonstrate logical-gate error suppression with increasing code distance. Calibrated circuit-level simulations in the supplementary materials indicate that, if the elementary physical error rates are reduced to approximately half of their present values, the H, S and CNOT schedules enter a regime with pronounced logical-error suppression as distance increases. Reaching this regime experimentally will require lower physical error rates and logical-operation schedules that preserve distance while minimizing spacetime cost\cite{hiraiNoMoreHooks2026,kishonySurfaceCodeOffTheHook2026,liaoDesignAutomationSpacetime2026}. The primitive layer demonstrated here can then be combined with state injection to prepare high-fidelity encoded magic states, either through cultivation or through distillation\cite{bravyi2005universal,gidneyMagicStateCultivation2024,rosenfeldMagicStateCultivation2025}. Together with scalable real-time decoding and feed-forward control, these elements provide a route from the present logical-instruction prototype to a universal fault-tolerant computing architecture in superconducting hardware.

\begin{acknowledgments}

\textbf{Funding:} 
This research was supported by the Innovation Program for Quantum Science and Technology-National Science and Technology Major Project (Grant No.~2021ZD0300200), National Natural Science Foundation of China (Grant No.~92476203), Anhui Initiative in Quantum Information Technologies, the Special funds from Jinan Science and Technology Bureau and Jinan high tech Zone Management Committee, Shanghai Municipal Science and Technology Major Project (Grant No.~2019SHZDZX01), and the XPLORER PRIZE.

\end{acknowledgments}

\clearpage
\onecolumngrid   
\setcounter{figure}{0}\renewcommand{\thefigure}{S\arabic{figure}}\renewcommand{\theHfigure}{suppfigure.\arabic{figure}}
\setcounter{table}{0}\renewcommand{\thetable}{S\arabic{table}}\renewcommand{\theHtable}{supptable.\arabic{table}}
\setcounter{equation}{0}\renewcommand{\theequation}{S\arabic{equation}}\renewcommand{\theHequation}{suppequation.\arabic{equation}}

\appendix

\begin{center}\large\textbf{Supplemental Material}\end{center}

\section{Appendix A: Experimental System Setup}
\subsection{QPU Introduction}
The experiments reported in this work were performed on the \textit{Zuchongzhi} 3.2 superconducting quantum processor~\cite{supp_he_experimental_2025}, a 107-qubit device featuring frequency-tunable transmon qubits with tunable nearest-neighbor couplers in a square lattice. Detailed descriptions of the processor design, fabrication, control electronics, and the general calibration framework---including single-qubit gates, two-qubit CZ gates, readout, and the all-microwave leakage reduction unit (LRU) and reset (RST) protocols for leakage suppression---are provided in Ref.~\cite{supp_he_experimental_2025} and the earlier characterizations of the same processor family in Refs.~\cite{supp_gao2025establishing, supp_jiang2026one}. Unless explicitly stated otherwise, all control and calibration procedures in this work follow those references.

\begin{figure*}[!htbp]
    \centering
    \includegraphics[width=0.89\textwidth]{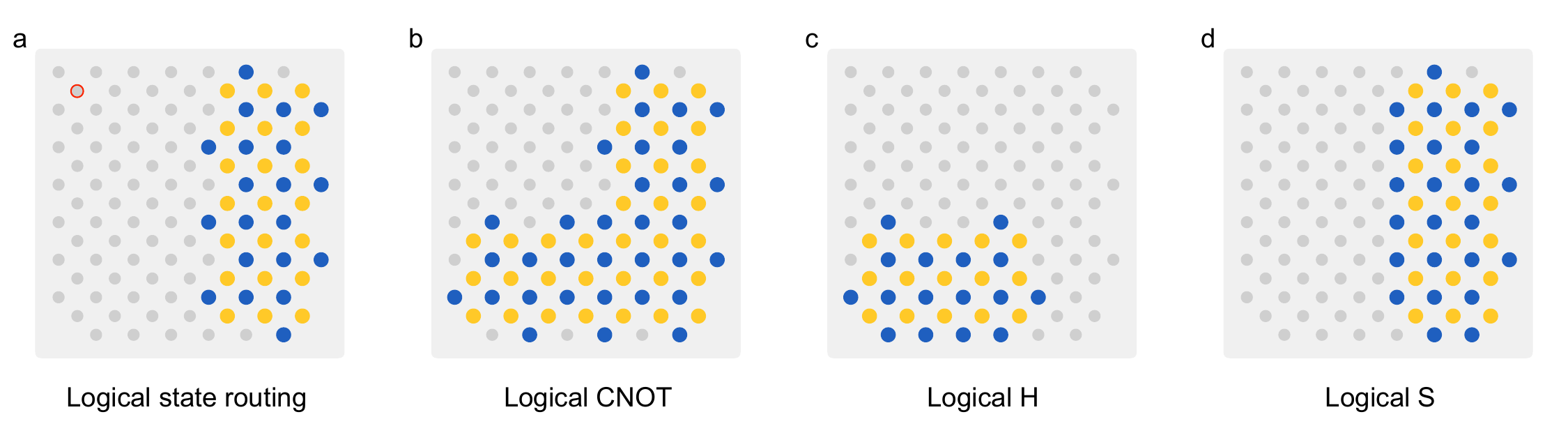}
    \caption{\textbf{Qubit layouts of the logical operations.} 
        Patch geometries on the 107-qubit \textit{Zuchongzhi} 3.2 processor for (a) logical state routing, (b) the logical CNOT gate, (c) the Hadamard ($H$) gate, and (d) the $S$ gate. Yellow and blue circles denote data and ancilla qubits, respectively. Light-gray circles indicate the remaining processor qubits not involved in this work. In (a), the red ring marks the qubit taken as the origin $(0,0)$ of the patch-coordinate labeling.}
    \label{fig:logical_layouts}
\end{figure*}

We employed the distance-3 rotated XZZX surface code~\cite{supp_bonilla2021xzzx} as the logical building block and implemented logical state routing together with a Clifford-generating set of logical gates---the Hadamard ($H$), phase ($S$), and controlled-NOT (CNOT) gates. Each operation occupies a sub-region of the processor defined by the corresponding patch geometry and routing ancillas, with the qubit layouts shown in Fig.~\ref{fig:logical_layouts} and the protocols detailed in the later sections.

All compiled circuits begin with active initialization on the active qubits used in the initial patch geometry. During each stabilizer-measurement cycle, LRUs are applied to data qubits and RST is applied to ancilla qubits, following the leakage-suppression protocol of Ref.~\cite{supp_he_experimental_2025}. When lattice-surgery or code-deformation operations introduce additional data or ancilla qubits, those qubits are reset before they become active in the code. This convention is used consistently in the memory, movement, CNOT-gate, $H$-gate, and $S$-gate experiments described below.

\subsection{Device Performance}

Among the logical operations implemented in this work, the logical CNOT gate uses the largest active sub-region of the processor. This pattern contains $65$ active qubits and $104$ active couplers, covering the majority of the qubits and couplers used by the logical movement, $H$-gate, and $S$-gate experiments. We therefore use the patterns in the logical CNOT-gate experiment as a representative physical-layer benchmark for the logical-operation experiments.

\begin{figure}[!htbp]
    \centering
    \includegraphics[width=0.4\textwidth]{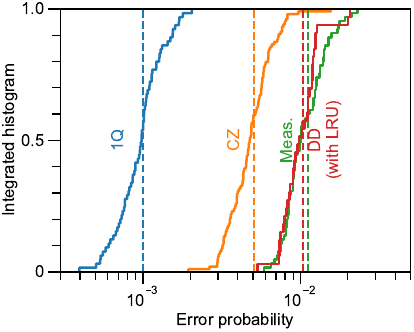}
    \caption{\textbf{Device performance.} 
        Cumulative distribution functions of four physical error metrics measured under parallel-gate operating conditions matching the logical CNOT protocol: single-qubit gate error (1Q, blue), two-qubit CZ-gate error (CZ, orange), readout assignment error (Meas., green), and dynamical-decoupling error with the leakage-reduction unit applied [DD (with LRU), red]. Vertical dotted lines mark the average of each distribution.}
    \label{performance_plot}
\end{figure}

Fig.~\ref{performance_plot} shows the cumulative distribution functions of the four principal physical error metrics: single-qubit gate error (1Q), two-qubit CZ gate error, readout assignment error (Meas.), and dynamical-decoupling error with LRU applied [DD (with LRU)]. The 1Q and CZ errors are calibrated by parallel cross-entropy benchmarking (XEB)~\cite{supp_arute2019quantum}; Readout errors are extracted from the $\ket{0}$ and $\ket{1}$ preparation-and-measurement sequence; the DD (with LRU) errors are characterized by interleaved randomized benchmarking, following Ref.~\cite{supp_he_experimental_2025}. 

The average 1Q, CZ, readout, and DD (with LRU) errors are $1.01 \times 10^{-3}$, $5.06 \times 10^{-3}$, $1.11 \times 10^{-2}$, $1.04 \times 10^{-2}$, respectively. These physical error rates provide the physical-layer baseline against which the performance of all logical operations reported in this work should be interpreted.

\section{Appendix B: Logical memory}

\subsection{Distance-3 memory baseline}
To characterize representative memory performance across the processor region used by the logical-operation experiments, we select five distance-3 rotated surface-code sub-patches that cover this region with minimal overlap.

For these memory experiments, the logical qubit on each selected sub-patch is prepared and measured in both the $X$ and $Z$ bases. In each basis, we measure both logical eigenstates to average over state-preparation and energy-relaxation asymmetries. The settings labeled \textit{Flip=False} and \textit{Flip=True} correspond to $\ket{+}$ and $\ket{-}$ in the $X$ basis, and to $\ket{0}$ and $\ket{1}$ in the $Z$ basis, respectively. The syndromes are decoded with the Tesseract decoder~\cite{supp_grbic2026accelerating}. The logical error rate (LER) per cycle, $\varepsilon$, is extracted by fitting the decoded logical fidelity, $F(r)$, as a function of the number of quantum error correction (QEC) cycles, $r$, to
\begin{equation}
    F(r) = F_0\,(1 - 2\varepsilon)^{r}
    \label{eq:fidelity_fit}
\end{equation}
where $F_0$ absorbs state preparation and measurement (SPAM) errors.

The extracted LERs are summarized in Table~\ref{tab:d3_baseline}. Across the five sub-patches, the LERs lie in the range $0.0128$ -- $0.0176$ in the $X$ basis and $0.0133$ -- $0.0197$ in the $Z$ basis. These results serve as a distance-3 memory reference for the subsequent logical operations experiments.

\begin{table}[!htbp]
    \centering
    \caption{\textbf{Logical memory performance for five distance-3 surface-code sub-patches.} 
        LER per cycle for five distance-3 sub-patches, measured in the $X$ and $Z$ bases for both logical eigenstates. The two flip settings correspond to $\ket{+}/\ket{-}$ in the $X$ basis and $\ket{0}/\ket{1}$ in the $Z$ basis. Each sub-patch is labeled by the coordinate $(x,y)$ of its upper-left data qubit, where $x$ ($y$) is the horizontal (vertical) displacement relative to the origin qubit shown in Fig.~\ref{fig:logical_layouts}a.}
    \label{tab:d3_baseline}
    \renewcommand{\arraystretch}{1.4}
    \begin{tabular}{|c|c|c|c|}
        \hline
        Coordinate & Flip & $X$-basis error rate & $Z$-basis error rate \\
        \hline
        \multirow{2}{*}{(0,4)} & \textit{False} & 0.0134(4) & 0.0150(5) \\
                               & \textit{True}  & 0.0128(5) & 0.0153(5) \\
        \hline
        \multirow{2}{*}{(2,4)} & \textit{False} & 0.0161(5) & 0.0181(6) \\
                               & \textit{True}  & 0.0169(5) & 0.0177(6) \\
        \hline
        \multirow{2}{*}{(4,0)} & \textit{False} & 0.0133(4) & 0.0145(5) \\
                               & \textit{True}  & 0.0131(5) & 0.0140(5) \\
        \hline
        \multirow{2}{*}{(4,2)} & \textit{False} & 0.0144(5) & 0.0139(4) \\
                               & \textit{True}  & 0.0149(5) & 0.0133(5) \\
        \hline
        \multirow{2}{*}{(4,4)} & \textit{False} & 0.0176(6) & 0.0190(6) \\
                               & \textit{True}  & 0.0174(6) & 0.0197(6) \\
        \hline
    \end{tabular}
\end{table}

\subsection{rectangular surface-code memory}

\begin{figure}[!htbp]
    \centering
    \includegraphics[width=0.5\textwidth]{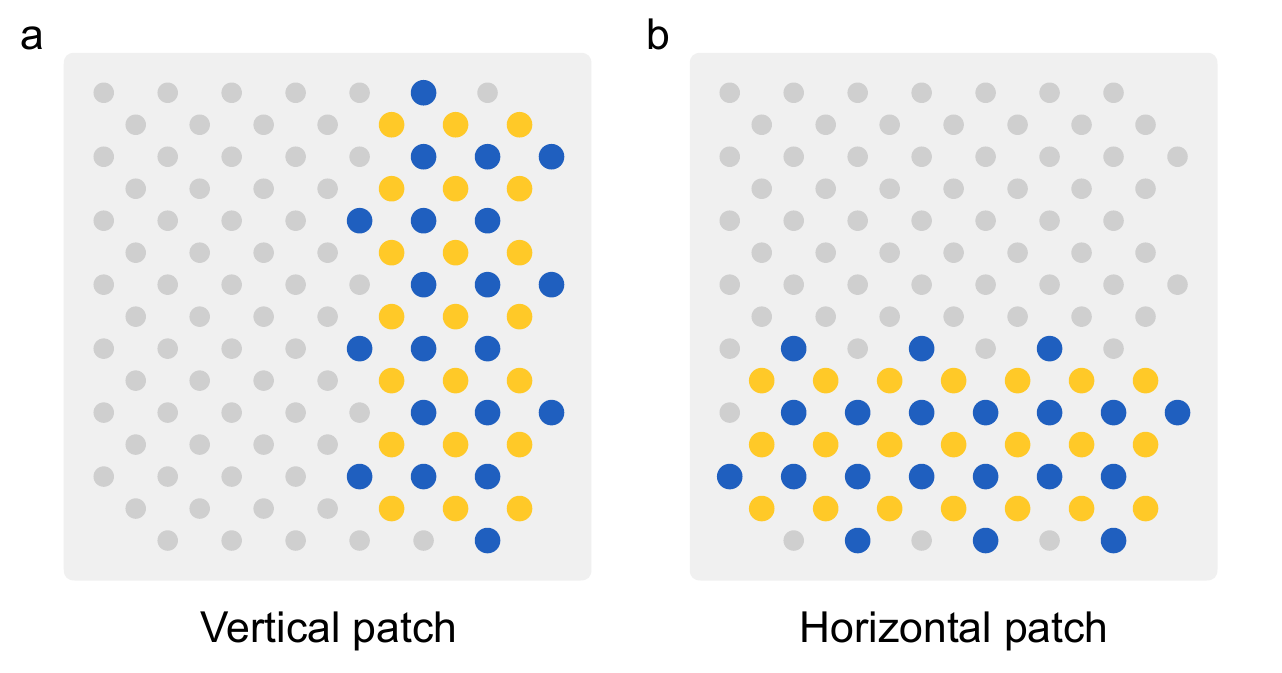}
    \caption{\textbf{Qubit layouts of the $3\times7$ surface-code patches.} 
        Locations of the vertical (a) and horizontal (b) $3\times7$ surface-code patches on the 107-qubit \textit{Zuchongzhi} 3.2 processor. Yellow and blue circles denote data and ancilla qubits, respectively; light-gray circles indicate processor qubits not used in the experiment. In both orientations, the top and bottom edges are $X$-type boundaries and the left and right edges are $Z$-type boundaries.}
    \label{fig:layout_3_7}
\end{figure}

The $3\times7$ rectangular surface-code patch is the intermediate code geometry used repeatedly in the logical-operation protocols. To characterize this memory block directly, we run repeated stabilizer-measurement cycles on $3\times7$ patches in the two spatial orientations shown in Fig.~\ref{fig:layout_3_7}, referred to as the vertical and horizontal patches. 

\begin{figure}[!htbp]
    \centering
    \includegraphics[width=0.53\textwidth]{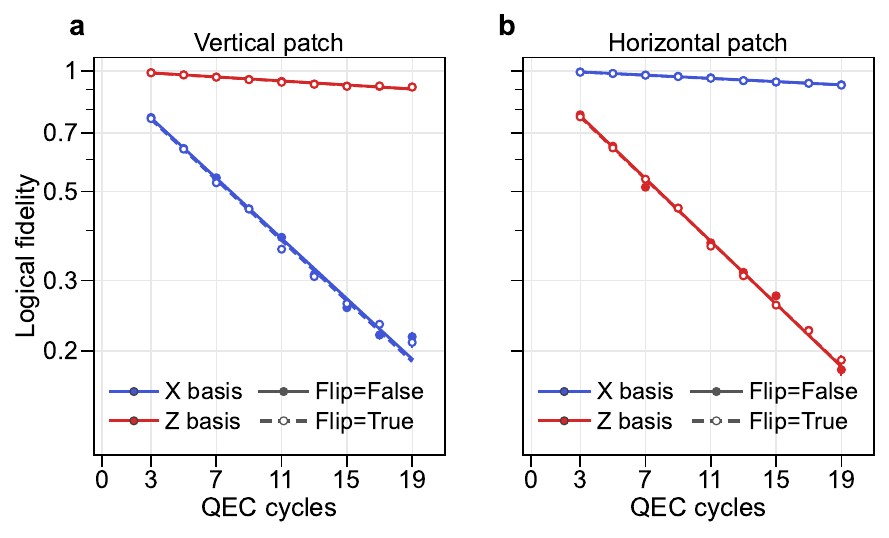}
    \caption{\textbf{Logical fidelity of the $3\times7$ surface-code memory.} 
        Decoded logical fidelity as a function of the number of QEC cycles for (a) the vertical patch and (b) the horizontal patch. Blue and red denote the $X$- and $Z$-basis memory experiments, respectively. Filled markers with solid lines correspond to \textit{Flip=False}, and open markers with dashed lines correspond to \textit{Flip=True}. Lines are exponential fits used to extract the logical error rate (LER) per cycle. Error bars denote $95\%$ confidence intervals from binomial tests and are smaller than the marker size.}
    \label{fig:fid_3_7}
\end{figure}

Fig.~\ref{fig:fid_3_7} shows the decoded logical fidelity as a function of the number of QEC cycles. The decay is strongly basis dependent. On the vertical patch, the $Z$-basis logical fidelity decays much more slowly than the $X$-basis fidelity, whereas the horizontal patch shows the opposite behavior. This anisotropy is expected from the rectangular geometry: the vertical patch has an effective distance of $7$ for preserving the $Z$-basis logical information and distance of $3$ for the orthogonal basis, while the horizontal patch interchanges these roles. The two flip settings nearly overlap in all cases, indicating that the memory performance is dominated by basis-dependent code distance rather than by the choice of logical eigenstate.

The extracted LERs are summarized in Table~\ref{tab:mem_3_7}. 
The basis with effective distance $7$ yields LERs of $2.88(6)\times10^{-3}$ to $2.89(6)\times10^{-3}$ on the vertical patch ($Z$ basis) and $2.31(5)\times10^{-3}$ to $2.32(5)\times10^{-3}$ on the horizontal patch ($X$ basis), where the number in parentheses denotes the $95\%$ confidence interval of a binomial test in this section. In the orthogonal basis, where the effective distance is $3$, the LERs are approximately $4.1\times10^{-2}$ -- $4.3\times10^{-2}$. Thus, the memory experiments on the $3\times7$ patches quantify the basis-dependent protection used as a reference for the logical movement and the logical operations in the following appendices.

\begin{table}[!htbp]
    \centering
    \caption{\textbf{Memory LER per cycle on the $3\times7$ surface-code patch.} 
        LER per cycle for the vertical and horizontal $3\times7$ patches, measured in the $X$ and $Z$ bases for both logical eigenstates. The effective distance is $7$ in the $Z$ basis on the vertical patch and in the $X$ basis on the horizontal patch.}
    \label{tab:mem_3_7}
    \renewcommand{\arraystretch}{1.4}
    \begin{tabular}{|c|c|c|c|}
        \hline
        Patch orientation & Flip & $X$-basis error rate & $Z$-basis error rate \\
        \hline
        \multirow{2}{*}{vertical} & \textit{False} & 0.04148(43) & 0.00289(6) \\
                                  & \textit{True}  & 0.04168(43) & 0.00288(6) \\
        \hline
        \multirow{2}{*}{horizontal} & \textit{False} & 0.00232(5) & 0.04312(43) \\
                                    & \textit{True}  & 0.00231(5) & 0.04280(44) \\
        \hline
    \end{tabular}
\end{table}

\begin{figure}[!htbp]
    \centering
    \includegraphics[width=0.60\textwidth]{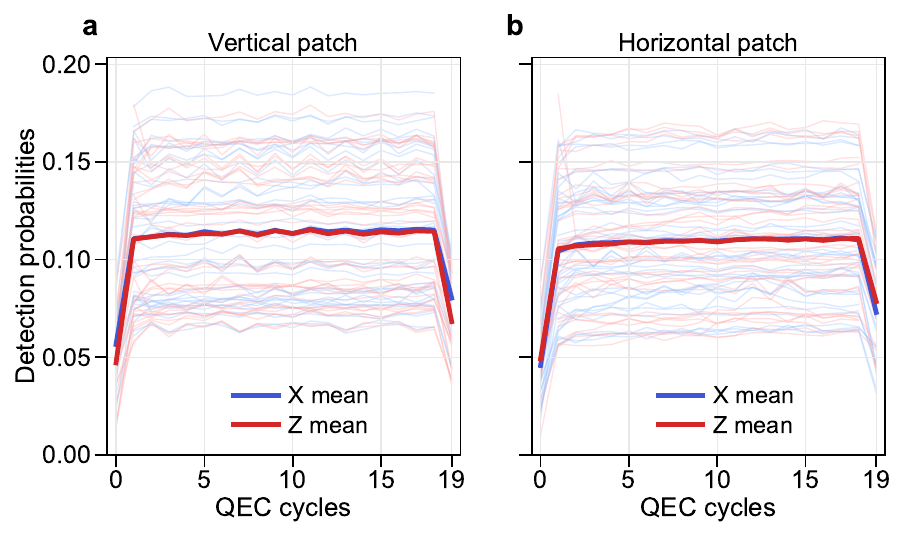}
    \caption{\textbf{Detection probabilities of the $3\times7$ surface-code memory.} 
        Detection probability as a function of the number of QEC cycles for (a) the vertical patch and (b) the horizontal patch. Blue and red correspond to the $X$- and $Z$-basis memory experiments, respectively. Light traces show individual-detector detection probabilities for both flip settings; thick traces show averages over detectors.}
    \label{fig:def_3_7}
\end{figure}

The detection probabilities provide a check that this basis anisotropy does not arise from substantially different physical error rates in the two orientations. Fig.~\ref{fig:def_3_7} shows the detection probability of each individual detector, together with the average over detectors for each basis. After the first syndrome-extraction cycle, the detection probabilities rise to a stable bulk plateau; the decrease in the final cycle is a boundary effect from terminating the syndrome-extraction sequence. The comparable average detection probabilities for the vertical and horizontal patches indicate that the large difference between the slow- and fast- logical-fidelity decay channels in Fig.~\ref{fig:fid_3_7} is primarily set by the basis-dependent effective distance, rather than by a substantial difference in the underlying physical error rates.

\section{Appendix C: Code-deformation and lattice-surgery primitives}

\begin{figure*}[t]
    \centering
    \includegraphics[width=\textwidth]{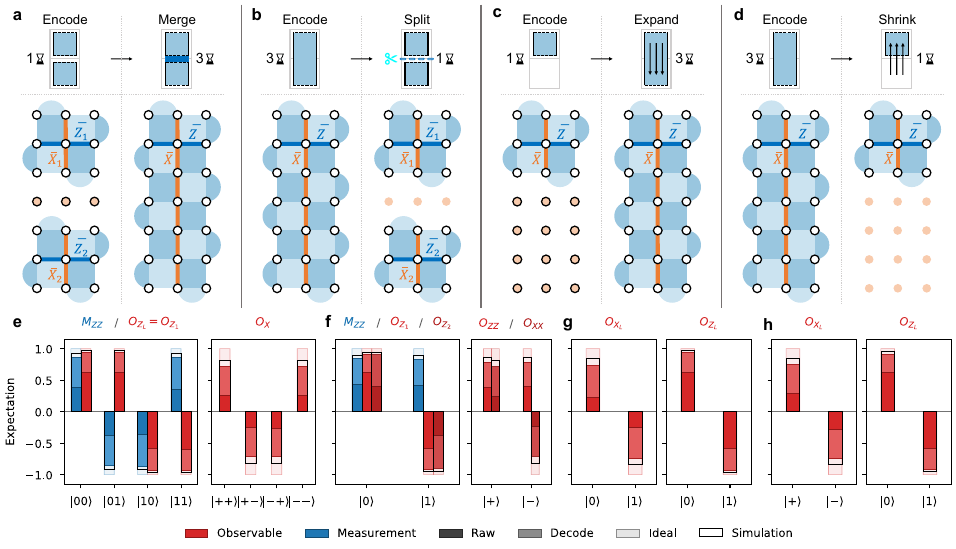}
    \caption{\textbf{Experimental benchmarking of code-deformation and lattice-surgery primitives.}
        \textbf{a--d,} Schematic operation and physical layout for merge, split, expansion and shrinkage, respectively.
        \textbf{e--h,} Raw and decoded expectation values for the corresponding operations, with ideal values and circuit-level simulations shown for comparison. Red bars denote output logical observables, and blue bars denote joint logical measurements. Dark filled, grey filled, pale filled and outlined bars denote raw data, decoded data, ideal values and circuit-level simulations, respectively. In the merge and split benchmarks, $M_{ZZ}$ is inferred from the interface stabilizer or measurement record. Expansion and shrinkage are assessed through the final logical observables.}
    \label{fig:primitive}
\end{figure*}

We benchmark the primitive operations that are reused in the logical movement and logical gate circuits: merge and split, and expansion and shrinkage~\cite{supp_horsman2012lattice,supp_litinskiGameSurfaceCodes2019}. Merge and split perform joint logical-parity measurements between neighbouring patches, whereas expansion and shrinkage change the footprint of one encoded patch while keeping the logical state encoded. In all four benchmarks, logical $X$- and $Z$-basis input states are prepared, the primitive operation is applied, and the relevant logical observables are decoded using the Tesseract decoder.

For the merge benchmark (Fig.~\ref{fig:primitive}a,e), two distance-$3$ patches are first encoded for one QEC cycle and then joined by activating the qubits and stabilizers in the intervening gap, forming a $3\times7$ rectangular surface-code patch. The interface stabilizer record gives the joint logical outcome $M_{ZZ}$, corresponding to the product of the two initial logical-$Z$ operators. The remaining logical observables are mapped as $Z_1\rightarrow Z_L$ and $X_1X_2\rightarrow X_L$, up to the measured Pauli frame. The three data qubits in the gap are initialized in the $X$ basis so that the newly introduced $X$-type stabilizers have deterministic initial values, and three QEC cycles are used to resolve the new stabilizer record.

The decoded observables in Fig.~\ref{fig:primitive}e verify this merge mapping. For the four $Z$-basis input states, the measured $M_{ZZ}$ signs agree with the expected input parity, and $O_{Z_L}=O_{Z_1}$ tracks the logical-$Z$ sign of the first input patch. For the four $X$-basis inputs, the decoded $O_X$ signs agree with the expected $X_1X_2\rightarrow X_L$ mapping. The observable magnitudes then show the basis-dependent error structure of the intermediate rectangular patch: the decoded $M_{ZZ}$ and $Z_L$ observables are larger than the $X_L$ observable, consistent with the asymmetric protection characterized in Appendix~B.

The split benchmark (Fig.~\ref{fig:primitive}b,f) implements the separation counterpart to merge. A $3\times7$ rectangular patch is run for three QEC cycles and is then separated into two distance-$3$ patches by deactivating the interface stabilizers and measuring the gap data qubits in the $X$ basis. This operation is the smooth-split primitive of lattice surgery~\cite{supp_horsman2012lattice}: up to Pauli-frame updates, it maps one encoded qubit into a two-patch correlated subspace,
\begin{equation}
    \alpha\ket{0}_L+\beta\ket{1}_L
    \longrightarrow
    \alpha\ket{00}_{12}+\beta\ket{11}_{12}.
    \label{eq:smooth_split}
\end{equation}
Thus split acts as a logical fanout in the $Z$ basis, rather than producing two independent copies. In the Heisenberg picture, the pre-split logical operator $Z_L$ can be represented by either output operator $Z_1$ or $Z_2$ within the fixed $Z_1Z_2$ parity sector, while $X_L$ maps to the joint operator $X_1X_2$. The split measurement record fixes the sign of the newly introduced $Z_1Z_2$ stabilizer and the associated Pauli frame.

The decoded observables in Fig.~\ref{fig:primitive}f verify this split mapping. For the $Z$-basis inputs, $O_{Z_1}$ and $O_{Z_2}$ carry the expected logical-$Z$ signs on the two output patches, and the measured $M_{ZZ}$ gives the expected relative parity. For the $X$-basis inputs, the decoded $O_{XX}$ and $O_{ZZ}$ observables have the expected signs for the two-patch Bell-state outputs generated by the split. 
The lower-bound of the Bell-state fidelity is
\begin{equation}
    F_{\rm Bell}^{(j)}
    \geq
    \frac{s_X^{(j)}\langle X_1X_2\rangle_j
    +s_Z^{(j)}\langle Z_1Z_2\rangle_j}{2},
\end{equation}
where $j$ labels the input state and $s_X^{(j)}$ and $s_Z^{(j)}$ are the ideal stabilizer signs. The minimum lower bound over the two Bell-state preparations is $0.748(4)$, above the separability bound of $0.5$, certifying entanglement between the two output logical patches. As in the merge benchmark, the larger decoded magnitudes of the $Z$-type observables compared with the $X_1X_2$ observable reflect the basis-dependent error structure of the rectangular intermediate patch.

The expansion and shrinkage benchmarks characterize code-deformation primitives that change the spatial footprint of a single logical patch. In expansion (Fig.~\ref{fig:primitive}c), a distance-$3$ patch is enlarged into the $3\times7$ rectangular patch by initializing the newly added data qubits in the $X$ basis and activating the additional stabilizers for three QEC cycles. This deformation preserves the encoded state while changing the available representatives of its logical operators. In the patch orientation used here, a logical-$X$ representative can extend along the long direction of the rectangular patch, while the logical-$Z$ representative is unchanged by the footprint extension. Consequently, the logical-$Z$ observable is protected by the distance-$7$ direction of the rectangular patch, whereas the logical-$X$ observable remains limited by the shorter distance-$3$ direction. Shrinkage (Fig.~\ref{fig:primitive}d) implements the reverse footprint change: the rectangular patch is reduced to a distance-$3$ patch by deactivating the auxiliary region and measuring the removed data qubits in the $X$ basis, with the measurement outcomes incorporated into the Pauli frame.

The decoded observables in Fig.~\ref{fig:primitive}g,h show the expected preservation of logical information through both footprint changes. In the expansion benchmark, the final $O_{Z_L}$ and $O_{X_L}$ observables have the expected signs after the patch is enlarged. In the shrinkage benchmark, the same sign agreement is observed after the auxiliary region is removed. The observable magnitudes then show the expected basis dependence: $Z_L$ observables are larger than $X_L$ observables for this patch orientation, matching the rectangular-memory characterization in Appendix~B. These expansion and shrinkage checks provide the code-deformation primitives used in the logical-movement and gate-construction experiments below.

\section{Appendix D: Decoding}

The decoding process converts the measured stabilizer records and final data-qubit readout into estimates of the logical observables used for state tomography, process tomography and logical-memory fits. We compare five decoders: the neural-network (NN) decoder, Tesseract~\cite{supp_grbic2026accelerating}, belief matching~\cite{supp_BeliefMatching_v010}, correlated matching~\cite{supp_he_experimental_2025} and PyMatching~\cite{supp_PyMatching_github}.

For every experimental shot, all measured records are retained; no post-selection is applied. We generate detector error models from calibrated circuit-level noise descriptions following the procedure used in Ref.~\cite{supp_he_experimental_2025}. These detector error models provide the common decoding prior for Tesseract and the matching-based decoders, and they also define the simulated data distribution used to pretrain the NN decoder.

\begin{table}[!ht]
    \centering
    \caption{\textbf{Pretraining and fine-tuning of the neural network decoder for each experimental configuration.}}
    \label{tab:training_schedule}
    \begin{tabular}{c c c c c}
        \hline
        Stage & Epoch & Learning rate & Batch size & Training samples \\
        \hline
        \multirow{3}{*}{Pretraining}
          & I   & $1\times10^{-4}$ & 512 & 8M \\
          & II  & $5\times10^{-5}$ & 512 & 8M \\
          & III & $1\times10^{-5}$ & 512 & 8M \\
        \hline
        \multirow{2}{*}{Fine-tuning}
          & I   & $1\times10^{-5}$ & 256 & 50,000 \\
          & II  & $5\times10^{-6}$ & 256 & 50,000 \\
        \hline
    \end{tabular}
\end{table}

\begin{figure}[H]
    \centering
    \includegraphics[width=0.5\textwidth]{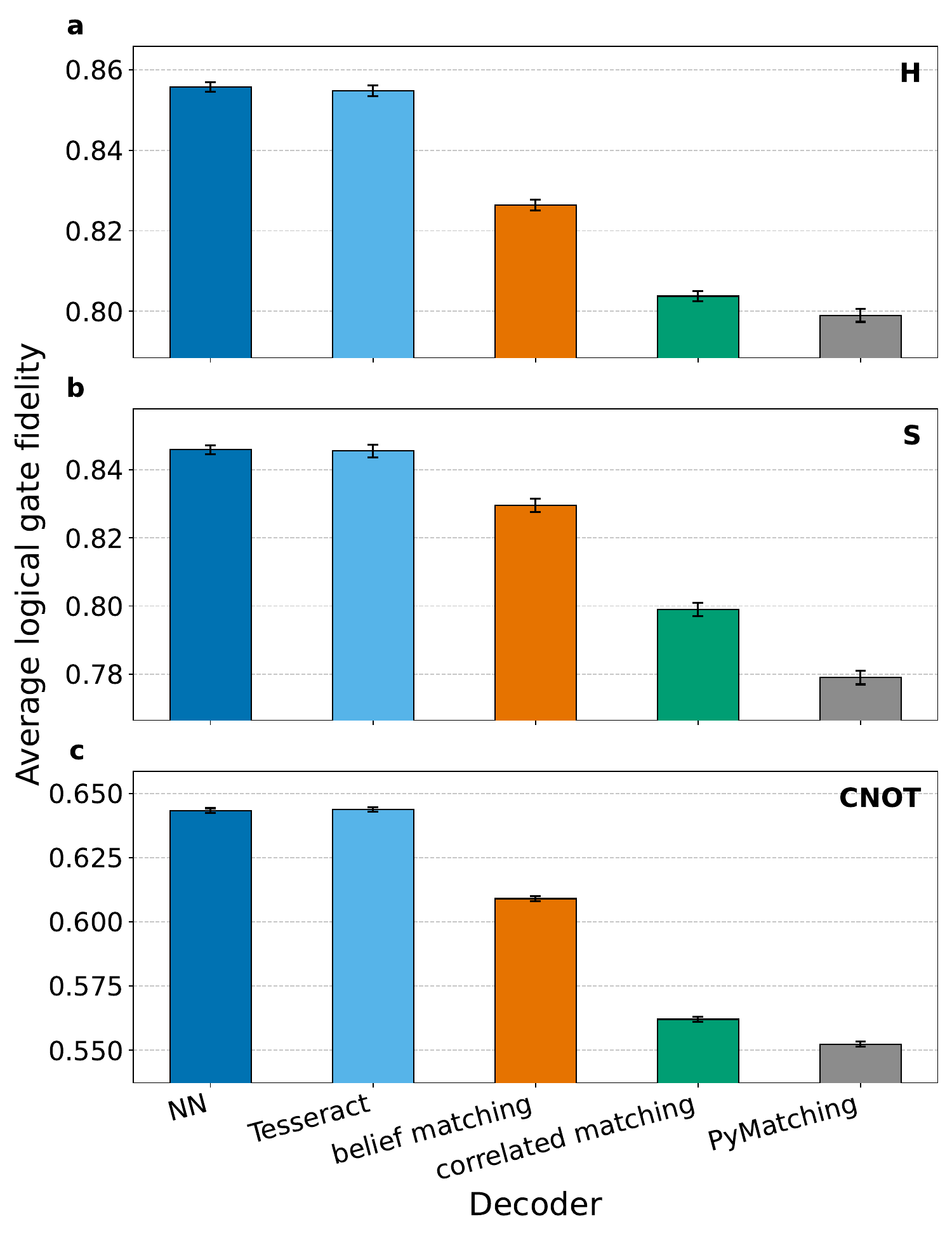}
    \caption{\textbf{Decoder comparison for logical-gate tomography.}
        \textbf{a--c,} Average logical gate fidelities reconstructed by quantum process tomography for the logical $H$ gate (\textbf{a}), $S$ gate (\textbf{b}) and CNOT gate (\textbf{c}). Bars compare five decoders applied to the same process-tomography data: the neural-network decoder, Tesseract, belief matching, correlated matching and PyMatching. Black error bars denote $95\%$ confidence intervals.}
    \label{figure_gate_fidelity}
\end{figure}

The neural network decoder shown in the main text is adapted from~\cite{supp_zhang2026learning}, which is similar to Google's AlphaQubit~\cite{supp_bausch2024learning} under global decoding. Since the demonstrated logical operation experiments are still relatively small in terms of spacetime volume, sliding-window-based parallelization is not yet necessary. Therefore, the model directly performs global prediction for each logical observable, making it convenient to fine-tune end-to-end on real-device data.

The training is conducted in two stages: pretraining on simulated data followed by fine-tuning on real-device data. As shown in Table~\ref{tab:training_schedule}, the pretraining stage uses 24M simulated samples in total, which are generated under a correlation-matrix-informed noise model for each experimental configuration, specified by the patch position, initialization state and measurement basis. After that, the pretrained model is fine-tuned end-to-end for two epochs using the even-indexed subset of the 100,000 shots real-device data. The remaining 50,000 shots (odd-indexed subset) are used to evaluate the decoding accuracy.

We benchmark the five decoders by applying them to the logical $H$, $S$ and CNOT process-tomography data and comparing the reconstructed average logical gate fidelities, as shown in Fig.~\ref{figure_gate_fidelity}. Across all three gates, the NN and Tesseract decoders give comparable fidelities and form the top-performing pair. Belief matching gives intermediate performance, while correlated matching and PyMatching yield lower reconstructed fidelities under the same data. Correlated matching nevertheless remains useful because it requires no training and has substantially lower decoding cost.

On the basis of this benchmark, we use the NN decoder for the main-text logical-operation data, where the most accurate decoded logical observables are required. We use Tesseract for the memory, primitive-operation and SPAM-calibration analyses in the supplementary materials, where it gives NN-comparable performance without experiment-specific fine-tuning. For the large distance-scaling simulations in Appendix~L, we use correlated matching because it avoids training and reduces decoding cost across the many simulated code distances and physical-error-rate scale factors.

\section{Appendix E: Logical process tomography}

We use logical process tomography to characterize the logical $H$, $S$ and $\mathrm{CNOT}$ gates in the main text~\cite{supp_chuangPrescriptionExperimentalDetermination1997}. For a Clifford gate, the ideal Pauli transfer matrix (PTM) has a sparse signed-permutation structure: each input Pauli operator is mapped to a single output Pauli operator, up to a sign. The reconstructed PTM therefore provides a direct check of the intended logical Pauli flow, while the overlap with the ideal process gives a compact tomography-based fidelity metric.

For each tomography circuit, all acquired shots are retained and no post-selection is applied. The decoded Pauli expectation values are assembled into the experimental PTM for both ideal-support elements and off-support elements whose ideal expectation value is zero. The latter do not increase the target-process overlap, but they serve as important diagnostics: large off-support values could signal coherent errors from improper calibration or spurious logical transfer rather than the intended Clifford action.

A key feature of the experimental logical tomography is that the reported PTMs include the state-preparation-and-measurement (SPAM) contribution of the tomography circuits. In particular, logical $Y$-basis SPAM operations are more costly than transversal $X$- and $Z$-basis SPAM, as quantified in Appendix~F. These basis-dependent SPAM errors are visible in the smaller PTM elements involving $Y$ operators, especially for the logical $H$ and $\mathrm{CNOT}$ gates where $Y$-basis access is used as an auxiliary tomography operation. We therefore report the experimentally reconstructed PTMs without projecting them onto the closest completely-positive trace-preserving(CPTP) channel. Instead, we use linear inversion to obtain the PTM and the corresponding process $\chi$ matrix, so that the displayed matrices preserve the directly observed basis-dependent tomography response.

Using the same process-matrix normalization for the reconstructed and ideal channels, we define the reconstructed process overlap as

\begin{equation}
    F_\text{process}=\text{Tr}(\chi_\text{exp}\chi_\text{ideal})
    \label{eq:processFidelity}
\end{equation}

The corresponding tomography-based average gate fidelity is then computed from
\begin{equation}
    F_\text{avg}=\frac{D F_\text{process}+1}{D+1},
    \label{eq:avgGateFidelity}
\end{equation}

where $D=2^n$ is the Hilbert-space dimension of the $n$-logical-qubit gate, i.e., $D=2$ for the single-qubit $H$ and $S$ gates and $D=4$ for the two-qubit $\mathrm{CNOT}$ gate. Because the experimental reconstruction is not SPAM-corrected, the quoted $F_\text{avg}$ values should be interpreted as SPAM-included tomographic fidelities of the implemented logical gate experiments.

The numerical simulations reported in the supplementary materials use a separate reconstruction convention. To reduce simulation cost, we simulate only deterministic logical observables, namely those whose ideal value is $\pm 1$ for the chosen input state, Clifford circuit and measurement basis. For example, for a given $\mathrm{CNOT}$ input state and readout basis, we sample only the output Pauli observable that is deterministic under the ideal Clifford evolution, while omitting observables whose ideal expectations are zero. The deterministic simulated expectation values are then combined using a maximum-likelihood-estimation(MLE)-based reconstruction constrained to a CPTP process matrix, and Eqs.~\ref{eq:processFidelity} and \ref{eq:avgGateFidelity} are used to obtain the plotted logical gate error rate. This MLE-based reconstruction is used for decoder-comparison and distance-scaling simulation metrics, whereas the experimental PTMs shown in the main text are obtained by linear inversion as described above.

\subsection{Post-selection considerations for logical-operation benchmarks}

Post-selection is a useful tool in quantum-error-correction experiments, but its interpretation depends on where it is applied in a fault-tolerant computation. For offline resource preparation, such as heralded ancilla preparation, magic-state preparation or distillation, rejected attempts can be discarded before the accepted resource is used in the computation~\cite{supp_lacroixScalingLogicColour2025,supp_rosenfeldMagicStateCultivation2025}. In that setting, the conditional fidelity is meaningful when reported together with the acceptance probability and the corresponding spacetime overhead. Post-selected analyses can also serve as diagnostics, because they show the performance obtained after conditioning on a specified syndrome or detection-event pattern, and related logical-operation experiments have reported such conditional metrics~\cite{supp_erhardEntanglingLogicalQubits2021a, supp_ryan-andersonHighfidelityTeleportationLogical2024, supp_bluvsteinFaulttolerantNeutralatomArchitecture2026, supp_besedinLatticeSurgeryRealized2026, supp_wangSuperconductingSurfacecodeProcessor2026}.

For algorithmic logical operations, the scalable benchmark is instead the unconditional decoded performance. Nontrivial syndrome records and detection events are expected during logical movement and logical gates; they are inputs to decoding and Pauli-frame tracking, not success criteria. Conditioning on a restricted set of records can increase the reported fidelity, but it converts a deterministic logical instruction into a conditional one whose acceptance probability generally decreases with circuit spacetime volume. The resulting conditional fidelity therefore does not represent the fidelity of a composable logical operation in a long fault-tolerant computation.

All logical-operation results in this work are reported without post-selection. The reported logical fidelities therefore quantify the unconditional, all-shot performance of composable logical operations under the stated decoder. Applying zero-detection-event post-selection to the same data would increase the conditional average gate fidelities to approximately $0.9952(10)$ for $H$, $0.9998(4)$ for $S$ and $0.9786(482)$ for CNOT, but with retained fractions of only $1.43(2)\%$, $1.44(2)\%$ and $0.02(1)\%$, respectively. These conditional values are diagnostic metrics rather than the primary scalable benchmark for the logical operations demonstrated here.

\section{Appendix F: Pauli basis state preparation and measurement}

Pauli-basis state preparation and measurement (SPAM) is a native component of surface-code computation. For a standard surface-code patch, logical $X$- and $Z$-basis SPAM can be implemented by transversal single-qubit preparation or readout, together with syndrome extraction and decoding. Access to the logical $Y$ basis is different: since $Y_L=iX_L Z_L$, a $Y$-basis eigenstate can't be prepared or measured by the same transversal data-qubit procedures without additional structure. Logical $Y$-basis SPAM therefore requires either a non-fault-tolerant state-injection/readout step or a fault-tolerant code deformation involving domain walls and twist defects~\cite{supp_gidney2024inplace}. This distinction is important for the logical gate experiments. The $Y$-basis settings used in logical movement, $H$-gate and CNOT-gate process tomography are auxiliary tomography operations, whereas fault-tolerant $Y$-basis access is an intrinsic part of the logical $S$-gate implementation.

\begin{figure*}
    \centering
    \includegraphics[width=1\columnwidth]{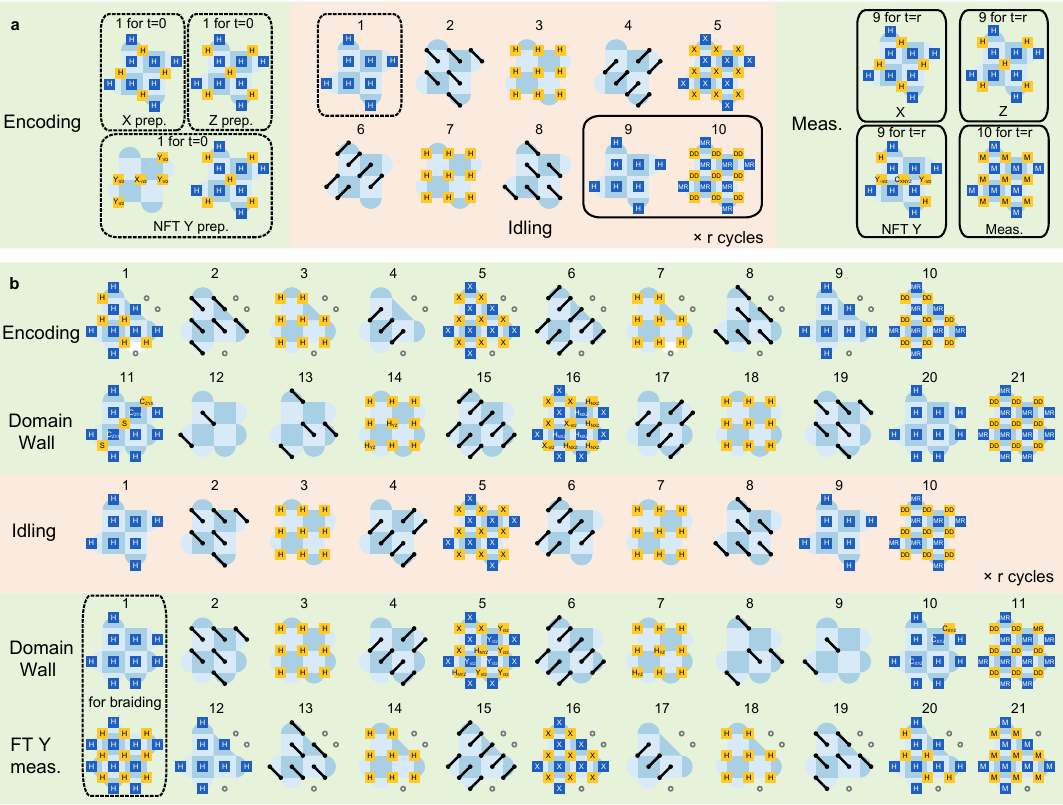}
    \caption{\textbf{Circuit schedules for Pauli-basis state preparation and measurement (SPAM) calibration.}
        \textbf{a,} Protocols for transversal $X$- and $Z$-basis SPAM and non-fault-tolerant (NFT) state-injection-based $Y$-basis SPAM. The left, middle and right blocks show logical-state preparation, $r$ inserted memory QEC cycles and matched logical-basis readout, respectively. \textbf{b,} Fault-tolerant (FT) $Y$-basis SPAM protocol based on diagonal twist-defect motion. The schedule consists of a fixed $Y$-basis preparation block, $r$ inserted memory QEC cycles and a fixed fault-tolerant $Y$-basis measurement block. The twist-braiding variant differs in the final domain-wall and measurement stages, where the gate pattern is rotated by $90^\circ$. Black links denote CZ gates, and site labels denote single-qubit gates, dynamical-decoupling operations, resets or measurements.}
    \label{fig:YSPAMcircuit}
\end{figure*}

To quantify these basis-dependent SPAM contributions, we benchmark four Pauli-basis SPAM protocols (Fig.~\ref{fig:YSPAMcircuit}): transversal fault-tolerant $X$-basis SPAM, transversal fault-tolerant $Z$-basis SPAM, non-fault-tolerant state-injection-based $Y$-basis SPAM, and fault-tolerant deformation-based $Y$-basis SPAM.
In each benchmark, a logical Pauli eigenstate is prepared, followed by $r=0$ to $6$ inserted memory QEC cycles and measurement in the same basis.
For the fault-tolerant $Y$-basis protocol, the cycle number $r$ denotes only the memory QEC cycles inserted between the fixed $Y$-basis preparation and measurement blocks.
The two preparation cycles and two measurement cycles required for fault-tolerant $Y$-basis access are included in every data point, so the $r=0$ point for this protocol already contains this four-cycle access overhead.
The plotted state fidelity is computed from the decoded logical Pauli expectation value as $F=(1+s\langle P_L\rangle_{\rm dec})/2$, where $P_L\in\{X_L,Y_L,Z_L\}$ is the target logical Pauli operator and $s=\pm1$ is the eigenvalue of the prepared state.

\begin{figure}
    \centering
    \includegraphics[width=0.45\textwidth]{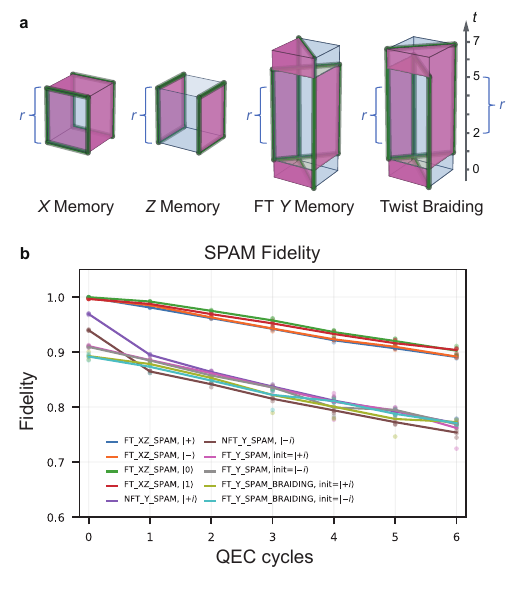}
    \caption{\textbf{Experimental calibration of Pauli-basis state preparation and measurement (SPAM).}
        \textbf{a,} Spacetime diagrams for the transversal $X$- and $Z$-basis memory protocols, the fault-tolerant $Y$-basis memory protocol and the twist-braiding $Y$-basis protocol. The non-fault-tolerant state-injection-based $Y$-basis protocol is not shown in this spacetime sketch. In each diagram, $r$ denotes the number of inserted memory QEC cycles; the fault-tolerant $Y$-basis protocols include fixed preparation and measurement overhead in addition to these cycles. \textbf{b,} Decoded logical state fidelity as a function of inserted QEC cycles for transversal $X/Z$-basis SPAM, state-injection-based $Y$-basis SPAM, fault-tolerant $Y$-basis SPAM and the twist-braiding variant. Colours distinguish the protocols and prepared logical eigenstates. Error bars denote $95\%$ confidence-interval half-widths.}
    \label{fig:YSPAM}
\end{figure}

The resulting fidelities are shown in Fig.~\ref{fig:YSPAM}b. Transversal $X$- and $Z$-basis SPAM has the highest fidelity throughout the sweep, with the sign-averaged fidelity decreasing from about $0.998$ at $r=0$ to about $0.897$ at $r=6$. Non-fault-tolerant $Y$-basis state injection has a higher zero-cycle fidelity than the fault-tolerant $Y$-basis protocol because it uses a shorter circuit, but it shows a sharper initial drop after the first QEC cycle. By contrast, the fault-tolerant $Y$-basis protocol starts with a lower fidelity because its fixed access overhead is already included at $r=0$, but its decay is smoother over the added memory cycles. At the present code distance and physical error rates, the non-fault-tolerant and fault-tolerant $Y$-basis protocols become comparable after the first few cycles. However, as physical error rates decrease and the code distance scales up, the fault-tolerant Y-basis method is expected to outperform the non-FT approach. 

To separate fixed SPAM overhead from memory-cycle decay, we fit the sign-averaged fidelity curves to the phenomenological form
\begin{equation}
    F_\text{state}(r)=\frac{F_0(1-2\epsilon)^r+1}{2} ,
\end{equation}
where $F_0$ is the fitted zero-cycle amplitude of $2F_\text{state}(r)-1$, corresponding to the excess fidelity above the random-output baseline extrapolated to $r=0$, and $\epsilon$ is the fitted LER per inserted QEC cycle. This fit provides two complementary diagnostics. The reduced $F_0$ values of the $Y$-basis protocols quantify the fixed loss in this zero-cycle amplitude associated with non-transversal or deformation-based $Y$-basis access. The larger fitted $\epsilon$ values quantify the faster decay during the inserted memory cycles. Using the $r=2$--$6$ data points, the transversal $X/Z$ protocols give $\epsilon=0.0203(11)$ for the $X$-basis states and $\epsilon=0.0196(8)$ for the $Z$-basis states, whereas the non-fault-tolerant and fault-tolerant $Y$-basis protocols give $\epsilon=0.0357(8)$ and $\epsilon=0.0356(49)$, respectively. Thus, both $Y$-basis protocols decay at about $1.8$ times the rate of the transversal $X/Z$ protocols over the fitted range. This faster decay is expected because an encoded $Y$-basis eigenstate is sensitive to logical failures in either the $X$- or $Z$-type channel; under approximately balanced logical noise, this gives $\epsilon_Y\simeq \epsilon_X+\epsilon_Z\approx 2\epsilon_{X/Z}$~\cite{supp_google2025below}.

These calibrated SPAM data provide the basis for interpreting the logical tomography results in the main text. For logical movement, the $H$ gate and the CNOT gate, $Y$-basis access is used only as an auxiliary tomography operation, and the smaller PTM entries involving $Y$ operators are consistent with the measured attenuation caused by $Y$-basis SPAM. For the logical $S$ gate, by contrast, fault-tolerant access to a logical $\ket{+i}$ state is part of the gate-teleportation circuit itself, so errors in the $Y$-basis access contribute to the implemented logical operation rather than only to tomography.

We also compare the fault-tolerant $Y$-basis protocol with a twist-braiding variant proposed in Ref.~\cite{supp_gidney2024inplace}, where the twist defect is fused through a $90^\circ$ rotation during the $Y$-basis measurement. When the number of idle rounds between state preparation and measurement is insufficient, $r<(d-1)/2$, this variant can introduce additional timelike error chains during the braiding portion of the circuit. At the present distance $d=3$, however, the measured fidelities of the twist-braiding and fault-tolerant $Y$-memory protocols are similar within the spread of the data. The comparison indicates that the current experiment is dominated by the fixed overhead and per-cycle decay of $Y$-basis access. 

\section{Appendix G: Logical movement}

\subsection{Circuit construction}

\begin{figure*}
    \centering
    \includegraphics[width=\columnwidth]{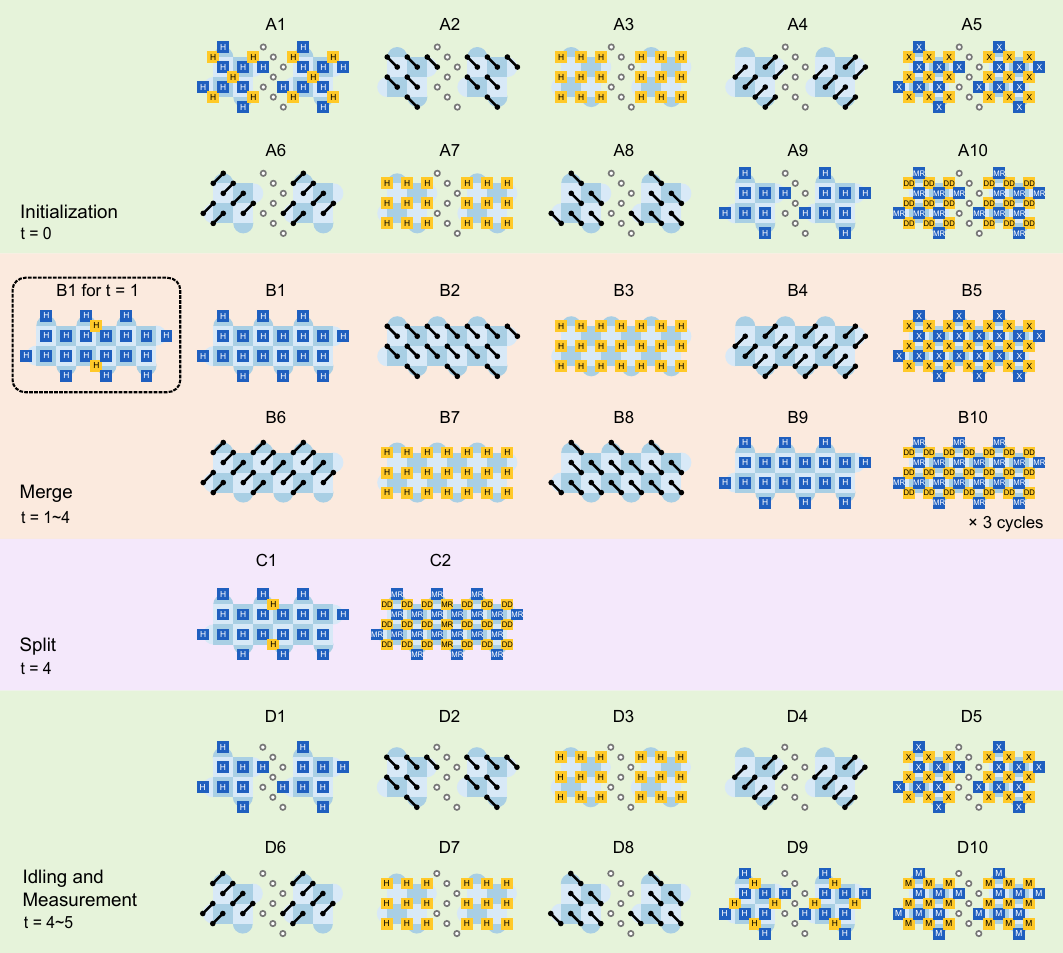}
    \caption{\textbf{Gate schedule for state teleportation by lattice surgery.}
        The schedule prepares an input patch and an auxiliary $\ket{+}_L$ patch, merges them into a $3\times7$ rectangular patch for three QEC cycles, splits the two patches, and measures the original input patch in the $X$ basis to complete state teleportation onto the auxiliary patch. The shaded bands mark the initialization, merge, split and final idling/readout stages. Lettered labels denote sequential gate layers; the dashed box marks the modified first merge layer at $t=1$. Black links denote CZ gates, and site labels denote single-qubit gates, dynamical-decoupling operations, resets or measurements.}
    \label{fig:CircuitTeleportation}
\end{figure*}

\begin{figure*}
    \centering
    \includegraphics[width=\columnwidth]{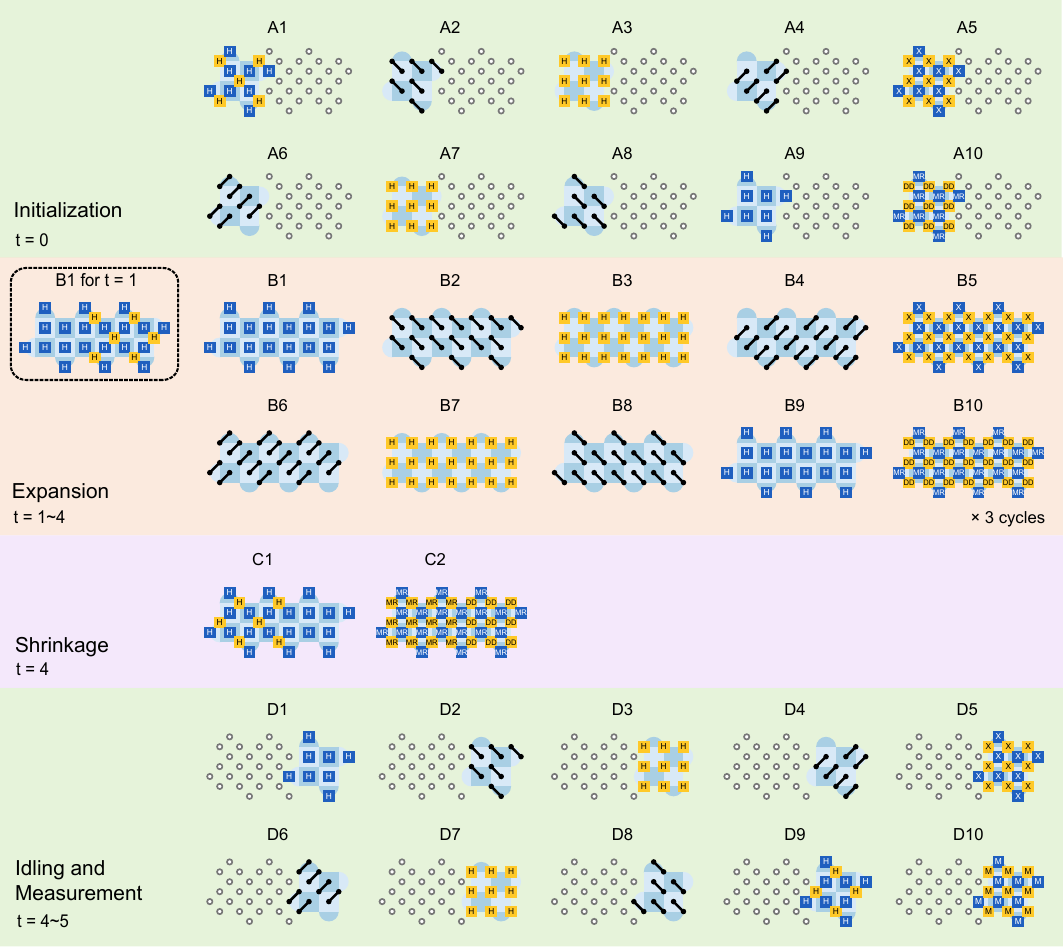}
    \caption{\textbf{Gate schedule for direct logical movement by code deformation.}
        The schedule expands the input patch into the target region, runs the resulting $3\times7$ rectangular patch for three QEC cycles, and shrinks the patch by measuring out the original region in the $X$ basis. The shaded bands mark the initialization, expansion, shrinkage and final idling/readout stages. Lettered labels denote sequential gate layers; the dashed box marks the modified first expansion layer at $t=1$. Black links denote CZ gates, and site labels denote single-qubit gates, dynamical-decoupling operations, resets or measurements.}
    \label{fig:CircuitMovement}
\end{figure*}

Logical state routing is the elementary operation that transfers encoded quantum information between different patch locations in a surface-code processor. In this work, logical state routing was implemented and compared using two closely related protocols: a lattice-surgery merge--split protocol and a direct expand--shrink protocol based on code deformation. Both protocols transfer the quantum information initially encoded in an input distance-$3$ patch to a neighbouring output region through an intermediate $3\times7$ rectangular surface-code geometry. The detailed gate schedules are shown in Fig.~\ref{fig:CircuitTeleportation} and Fig.~\ref{fig:CircuitMovement}.

In the merge--split protocol, the input patch is prepared in the target logical state and an auxiliary patch is prepared in $\ket{+}_L$. After one QEC cycle on the two separated patches, the patches are merged into a $3\times7$ rectangular surface-code patch. The newly activated stabilizer records determine the Pauli frame and allow the logical-$Z$ representative to be tracked from the input side to the auxiliary side. Three QEC cycles are then applied to the rectangular patch, followed by a split operation. Up to Pauli-frame corrections, this split maps the state into the correlated subspace $\alpha\ket{00}_{12}+\beta\ket{11}_{12}$. A final QEC cycle and an $X$-basis measurement of the original input patch complete the teleportation onto the auxiliary patch. The expand--shrink protocol removes the independent auxiliary-patch preparation and final input-patch QEC stage: the input patch is directly expanded into the target region, run for three QEC cycles as a rectangular patch, and then shrunk by measuring out the original region in the $X$ basis. Thus both protocols have the same displayed $1+3+1$ routing schedule, but not the same logical-observable spacetime volume. In the expand--shrink protocol, the encoded information is always stored in a single logical patch or in its expanded deformation. In the merge--split protocol, by contrast, the logical-$X$ observable passes through two additional distance-$3$ QEC cycles: the auxiliary-patch cycle before the merge and the input-patch cycle after the split. For general code distance $d$, the routing footprint scales as $d\times(2d+1)\times(d+2)$.

\subsection{Error analysis}

The decoded state fidelities for the four logical Pauli eigenstate inputs are shown in Fig.~2 of the main text. For the merge--split protocol, the fidelities are $0.908(2)$, $0.911(4)$, $0.841(2)$ and $0.842(4)$ for the input states $\ket{0}$, $\ket{1}$, $\ket{+}$ and $\ket{-}$, respectively. For the expand--shrink protocol, the corresponding fidelities are $0.908(3)$, $0.905(1)$, $0.860(1)$ and $0.863(3)$. Both protocols therefore preserve $Z$-basis inputs more strongly than $X$-basis inputs. This basis dependence is consistent with the rectangular-memory characterization in Appendix~B: during the rectangular movement stage, logical-$Z$ information is protected along the effective distance-$7$ direction of the rectangular patch, whereas logical-$X$ information remains limited by the effective distance-$3$ direction. The two protocols are nearly equivalent for $\ket{0}$ and $\ket{1}$, while the expand--shrink protocol improves the $\ket{+}$ and $\ket{-}$ fidelities by approximately $0.02$.

To identify the origin of this protocol-dependent difference, we compare the data with Pauli simulations and a simple LER product model. Independent asymmetric Pauli-noise simulations give $0.947(2)$ and $0.947(1)$ for the merge--split protocol in the $Z$ basis, and $0.944(1)$ and $0.944(2)$ for the expand--shrink protocol, indicating only a small difference between the two protocols. In the $X$ basis, the same simulations give $0.899(2)$ and $0.899(2)$ for the merge--split protocol, compared with $0.914(2)$ and $0.915(3)$ for the expand--shrink protocol. The slight simulated separation in the $X$ basis follows the observable-flow picture: the logical-$X$ operator in the merge--split protocol accumulates errors over the two additional distance-$3$ QEC cycles. Quantitatively, we model a logical circuit as a sequence of QEC cycles with cycle-dependent logical error rates $\varepsilon_i$. For each cycle, we define the logical Pauli fidelity as $F_i=1-2\varepsilon_i$. Under the assumption that different QEC cycles have independent fidelities, the circuit-level logical Pauli fidelity is estimated as

\begin{equation}
F=\prod_i F_i
\label{eq:ProductFidelity}
\end{equation}
The corresponding final Pauli-eigenstate fidelity is $F_{\rm state}=(1+F)/2$. Using the measured distance-$3$ memory LER and the short-direction LER of the $3\times7$ rectangular patch gives an estimated $X$-basis fidelity of approximately $0.860(1)$ for the expand--shrink protocol. Including the two additional distance-$3$ QEC cycles of the merge--split protocol lowers the estimate to approximately $0.838(2)$, consistent with the measured values. Here, the parenthetical uncertainties of the estimated values are calculated through standard propagation of uncertainty. In the $Z$ basis, however, the same estimate gives $0.962(1)$, well above the measured fidelities. This overestimate occurs because the model uses the distance-$7$ spatial logical fidelity of the rectangular patch, while the rectangular memory lasts only three QEC cycles and can still support timelike weight-$3$ logical error chains along the syndrome-time direction. Because this error mechanism is uniquely inherent to logical movement, it accounts for the discrepancy with the memory-based estimation.

These results show that the main basis dependence of logical state routing is set by the asymmetric effective distance of the rectangular intermediate patch, while the protocol-dependent $X$-basis difference is explained by the additional logical-$X$ spacetime volume in the merge--split protocol. The Pauli simulations reproduce this relative trend but overestimate the absolute fidelities, as expected for a simplified model that does not fully capture leakage, residual crosstalk and other device-specific non-Pauli errors. The expand--shrink protocol is therefore preferred in later logical-gate designs whenever it can replace an ancilla preparation or final measurement step without changing the intended logical operation, because it reduces the observable-flow volume of the relevant logical operators.

\section{Appendix H: Logical CNOT gate}

\subsection{Circuit construction}

The logical CNOT gate is implemented using the standard ancilla-mediated lattice-surgery identity. Let $C$, $T$ and $A$ denote the control, target and ancilla patches, respectively. The ancilla is prepared in $\ket{+}_A$, followed by a joint $Z_CZ_A$ measurement, a joint $X_AX_T$ measurement, and a final $Z_A$ measurement of the ancilla. In the Heisenberg picture, the intended logical action is 
$X_C\rightarrow X_CX_T$, $Z_T\rightarrow Z_CZ_T$, $Z_C\rightarrow Z_C$ and $X_T\rightarrow X_T$.
With measurement outcomes $m_{ZZ},m_{XX},m_Z\in\{\pm1\}$ for $Z_CZ_A$, $X_AX_T$ and $Z_A$, respectively, the byproducts are absorbed into the Pauli frame: the frame contains a $Z_C$ byproduct if $m_{XX}=-1$, and an $X_T$ byproduct if $m_{ZZ}m_Z=-1$. These updates are tracked classically in the decoded logical observables; no physical Pauli correction is applied during the experiment.

The compiled experimental schedule implements this lattice-surgery identity using the movement primitives characterized in Appendix~G. Instead of preparing an independent ancilla patch, we expand the control patch into the ancilla region. Similarly, instead of maintaining a separately encoded ancilla patch through final $Z$-basis readout, we incorporate the ancilla region into the target patch and shrink the rectangular patch back to the target region. These substitutions reduce the spacetime volume of the relevant observable flows. The switch between the control--ancilla split and the target--ancilla merge is implemented by data-qubit measurements and initializations, so no additional syndrome-extraction cycle is inserted between the two parity-measurement stages. We also omit extra boundary QEC cycles before the first expansion and after the final shrinkage because the control and target patches each undergo at least $d$ QEC cycles between logical state preparation and final readout, allowing measurement errors to be handled by the decoder.

\begin{figure*}
    \centering
    \includegraphics[width=0.85\textwidth]{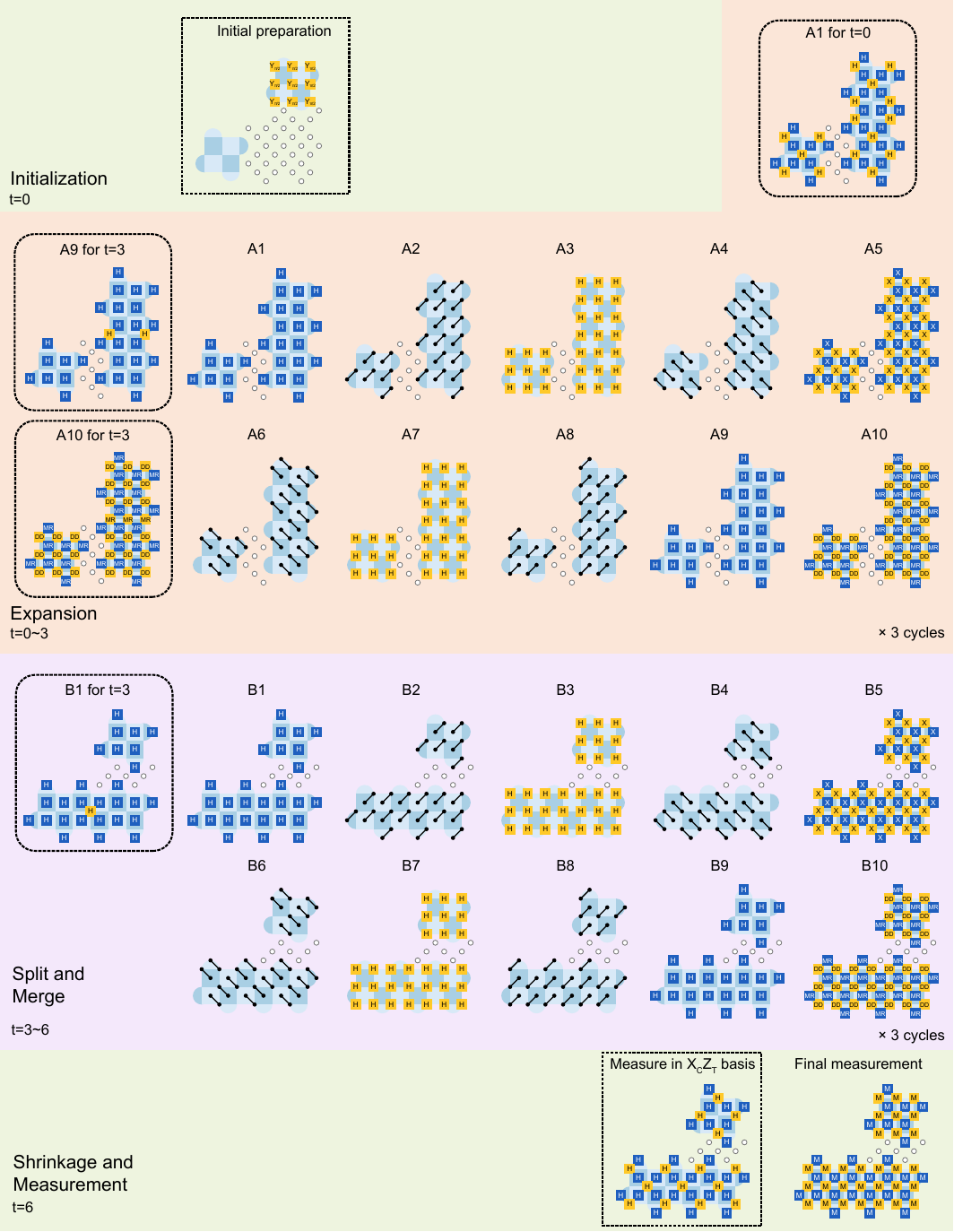}
    \caption{\textbf{Gate schedule for the logical CNOT circuit.}
        The distance-$3$ schedule is shown for the $\ket{+0}$ tomography configuration. The shaded bands mark initial preparation, expansion, split-and-merge, and shrinkage/readout stages. In the expansion stage, layers A1--A10 are repeated for three QEC cycles while the control patch is expanded into the ancilla region and the target patch idles. In the split-and-merge stage, layers B1--B10 are repeated for three QEC cycles while the ancilla coupling is transferred from the control side to the target side. Dashed boxes indicate modified boundary layers at stage transitions. Black links denote CZ gates, and site labels denote single-qubit gates, dynamical-decoupling operations, resets or measurements. The full experimental schedule uses a $7\times7$ data-qubit footprint and six syndrome-extraction cycles.}
    \label{fig:CircuitCNOT}
\end{figure*}

The resulting distance-$3$ gate sequence is shown in Fig.~\ref{fig:CircuitCNOT}. More generally, the first $d$ QEC cycles implement the $Z_CZ_A$ measurement through an expand--split stage involving the control and ancilla regions, while the target patch remains idle. The following $d$ QEC cycles implement the $X_AX_T$ measurement through a merge--shrink stage involving the ancilla and target regions, while the control patch remains idle. The maximum data-qubit footprint is a $(2d+1)\times(2d+1)$ square, and the depth is $2d$ syndrome-extraction cycles. We therefore count the spacetime volume of the compiled logical CNOT schedule as $(2d+1)\times(2d+1)\times 2d$. For the experimental distance-$3$ implementation, this corresponds to a $7\times7$ data-qubit footprint and six QEC cycles.

\subsection{Error analysis}

The reconstructed CNOT Pauli transfer matrix (PTM) provides a direct observable-flow diagnostic of the gate. Table~\ref{tab:CNOT_ptm_nonzero} lists the ideal-support PTM elements, together with the calibrated Pauli simulations with the same tomography-boundary SPAM convention as the experiment and the corresponding simulations with tomography SPAM removed. The signs follow the ideal CNOT Pauli action; in the discussion below we compare the magnitudes of the transfer amplitudes.

\begin{table*}[t]
    \centering
    \caption{\textbf{Ideal-support Pauli-transfer-matrix elements of the logical CNOT gate.}
        The table lists the non-zero PTM elements expected for an ideal CNOT, the experimentally reconstructed values, the calibrated Pauli-simulation values including tomography SPAM, and the corresponding simulation values with tomography SPAM removed. Parenthetical uncertainties are 95\% confidence-interval half widths, obtained from five decoding trials for the bootstraped experimental entries or repeated simulation.}
    \label{tab:CNOT_ptm_nonzero}
    \footnotesize
    \begin{ruledtabular}
    \begin{tabular}{cccccc}
        Input Pauli & Ideal output Pauli & Ideal & Experiment & Simulation & Sim. without SPAM \\
        \hline
        $II$ & $II$ & $1.0000$ & $1.0000$ & $1.0000$ & $1.0000$ \\
        $IX$ & $IX$ & $1.0000$ & $0.8635(24)$ & $0.8890(23)$ & $0.9060(89)$ \\
        $IY$ & $ZY$ & $1.0000$ & $0.4186(39)$ & $0.4770(53)$ & $0.6330(86)$ \\
        $IZ$ & $ZZ$ & $1.0000$ & $0.5821(29)$ & $0.6478(17)$ & $0.6839(41)$ \\
        $XI$ & $XX$ & $1.0000$ & $0.6197(56)$ & $0.7138(26)$ & $0.7463(53)$ \\
        $XX$ & $XI$ & $1.0000$ & $0.5674(107)$ & $0.6616(55)$ & $0.6949(98)$ \\
        $XY$ & $YZ$ & $1.0000$ & $0.2733(232)$ & $0.3427(83)$ & $0.5091(125)$ \\
        $XZ$ & $YY$ & $-1.0000$ & $-0.2973(63)$ & $-0.3609(55)$ & $-0.5379(117)$ \\
        $YI$ & $YX$ & $1.0000$ & $0.4514(81)$ & $0.5199(57)$ & $0.6972(91)$ \\
        $YX$ & $YI$ & $1.0000$ & $0.4090(69)$ & $0.4795(78)$ & $0.6480(75)$ \\
        $YY$ & $XZ$ & $-1.0000$ & $-0.3290(163)$ & $-0.4019(64)$ & $-0.5337(133)$ \\
        $YZ$ & $XY$ & $1.0000$ & $0.3360(31)$ & $0.4108(36)$ & $0.5668(128)$ \\
        $ZI$ & $ZI$ & $1.0000$ & $0.8619(11)$ & $0.8953(12)$ & $0.9140(37)$ \\
        $ZX$ & $ZX$ & $1.0000$ & $0.7429(86)$ & $0.7960(49)$ & $0.8310(107)$ \\
        $ZY$ & $IY$ & $1.0000$ & $0.4626(36)$ & $0.5178(37)$ & $0.6756(97)$ \\
        $ZZ$ & $IZ$ & $1.0000$ & $0.6408(24)$ & $0.7023(13)$ & $0.7291(57)$ \\
    \end{tabular}
    \end{ruledtabular}
\end{table*}

The experimental PTM elements separate into a small number of observable-flow classes. The invariant operators $ZI$ and $IX$ have the largest non-trivial amplitudes, $0.8619(11)$ and $0.8635(24)$, because their representatives remain on the stationary output patches and do not spread through the lattice-surgery part of the schedule. The product operator $ZX$ is then well approximated by the product of these two independent contrasts: $0.8619\times0.8635=0.7443$, consistent with the measured $ZX$ amplitude of $0.7429(86)$. Within the present uncertainty, this agreement indicates that no large correlated logical error between these two stationary observable flows is resolved.

The next two groups contain the non-trivial CNOT flows that pass through the rectangular-patch portion of the schedule. The $ZZ\rightarrow IZ$ and $XI\rightarrow XX$ elements have amplitudes $0.6408(24)$ and $0.6197(56)$. These observables undergo one expansion/shrinkage segment, one movement segment and accumulate errors over the rectangular patch together with an additional distance-$3$ part of the schedule. The $IZ\rightarrow ZZ$ and $XX\rightarrow XI$ elements are lower, $0.5821(29)$ and $0.5674(107)$, because their observable flows contain the same deformation structure plus a larger distance-$3$ spacetime volume. Thus the hierarchy among the non-$Y$ PTM elements is consistent with the spacetime volume occupied by the corresponding logical observables.

All remaining ideal-support elements involve at least one logical $Y$ operator. Their experimental amplitudes lie between $0.2733(232)$ and $0.4626(36)$, below the non-$Y$ groups. This reduction is expected because the logical $Y$ settings in the CNOT tomography are auxiliary tomography operations implemented through non-transversal $Y$-basis preparation and readout, rather than native transversal $X/Z$ surface-code SPAM. Appendix~F shows that these $Y$-basis SPAM operations have lower contrast and faster memory-cycle decay. Consistently, removing tomography-boundary SPAM in simulation raises the $Y$-containing elements much more strongly than the non-$Y$ elements: for example, $YI\rightarrow YX$ increases from $0.5199(57)$ to $0.6972(91)$, whereas $ZI\rightarrow ZI$ increases only from $0.8953(12)$ to $0.9140(37)$. The no-SPAM simulations nevertheless preserve a residual spread among the $Y$-containing elements, showing that SPAM is not the only effect; the remaining differences follow the spacetime volume of the underlying non-$Y$ observable components.

Overall, the experimental PTM, the calibrated Pauli simulations, and the SPAM-removed simulations support a two-part interpretation of the CNOT process-tomography data. The dominant ideal-support structure verifies the intended CNOT Pauli flow. The transfer-amplitude hierarchy is then explained by the observable-flow spacetime volume, with an additional attenuation for elements that require logical $Y$-basis tomography access.

The Bell-state experiments in main-text Fig.~3c provide a complementary, state-level test of the entangling action. For each input state $j$, the ideal CNOT output is a two-logical-qubit stabilizer state. Let $P_1^{(j)}$ and $P_2^{(j)}$ denote two signed stabilizer generators of this target state, with the ideal signs absorbed so that the target expectations are $+1$. The decoded Bell-state fidelity is calculated as
\begin{equation}
    F_{\rm Bell}^{(j)}
    = \frac{1+\langle P_1^{(j)}\rangle+\langle P_2^{(j)}\rangle+\langle P_1^{(j)}P_2^{(j)}\rangle}{4}.
    \label{eq:CNOTBellFidelity}
\end{equation}
For any pure Bell state, a fidelity greater than $1/2$ excludes all separable two-qubit states and therefore witnesses entanglement between the two output logical patches. The raw, undecoded fidelities are approximately $0.28$--$0.30$, whereas the decoded values are $0.6238(23)$, $0.6204(18)$, $0.5787(26)$, $0.5908(29)$, $0.5881(65)$ and $0.5503(14)$ for the $\ket{+0}$, $\ket{+1}$, $\ket{+i}$, $\ket{i0}$, $\ket{i1}$ and $\ket{ii}$ inputs, respectively. Thus all six decoded output states lie above the separability threshold, consistent with deterministic entanglement generation by the logical CNOT gate.

The input-state dependence of these Bell fidelities follows the same error hierarchy seen in the PTM. The $\ket{+0}$ and $\ket{+1}$ inputs give the largest decoded fidelities because their Bell-state stabilizer measurements contain the smallest contribution from $Y$-basis tomography and from the CNOT observable-flow volume. Inputs involving logical $\ket{i}$ require additional $Y$-basis preparation or readout. The $\ket{i0}$ and $\ket{i1}$ cases therefore have lower fidelities, and the $\ket{ii}$ case is lowest because all stabilizer terms used in Eq.~\ref{eq:CNOTBellFidelity} involve $Y$-basis access. The $\ket{+i}$ input also requires two $Y$-related stabilizer measurements, and its remaining non-$Y$ stabilizer occupies a larger CNOT spacetime volume than in the $\ket{i0}$ and $\ket{i1}$ cases, explaining its slightly lower fidelity.

Calibrated Pauli simulations reproduce this ordering, giving Bell fidelities of $0.6800(28)$, $0.6803(16)$, $0.6328(8)$, $0.6429(24)$, $0.6434(16)$ and $0.5987(7)$ for the same six input states. This agreement indicates that the observed asymmetry is not dominated by statistical fluctuation, but is captured by the measured circuit-level noise and the compiled CNOT schedule. When tomography-boundary SPAM is removed in simulation, the corresponding fidelities increase to $0.7439$, $0.7469$, $0.7205$, $0.7397$, $0.7380$ and $0.7114$, and the spread between different inputs is substantially reduced. The remaining spread follows the spacetime volume of the stabilizer observables used to evaluate each Bell state, while the larger SPAM-included spread is primarily caused by the lower contrast of logical $Y$-basis access.

A simple phenomenological estimate based on the PTM observable classes gives the same qualitative ordering. Using Eq.~\ref{eq:CNOTBellFidelity} with the class-dependent observable contrasts gives estimated Bell fidelities of approximately $0.632$ for the $\ket{+0}$ and $\ket{+1}$ outputs, $0.603$ for the $\ket{i0}$ and $\ket{i1}$ outputs, $0.592$ for the $\ket{+i}$ output, and $0.562$ for the $\ket{ii}$ output. These estimates are not a substitute for the full circuit-level simulation, but they make the dominant mechanism transparent: Bell-state fidelities are set by the decoded stabilizer contrasts, which are in turn controlled by the CNOT observable-flow volume and by whether the stabilizer measurement requires logical $Y$-basis tomography access.

\section{Appendix I: Logical H gate}

\subsection{Circuit construction}

The logical Hadamard gate exchanges the logical Pauli axes, $X_L\leftrightarrow Z_L$, and changes the sign of $Y_L$. In a rotated surface-code patch, a transversal physical-H layer exchanges $X$- and $Z$-type stabilizers and therefore also exchanges the boundary types. A standalone transversal-H layer consequently leaves the output encoded in a rotated boundary frame. We implement the logical $H$ gate by combining a geometric $90^\circ$ rotation of the patch boundary frame with a final Hadamard domain wall, so that the output logical patch has the intended orientation while the Pauli flow realizes the Hadamard transformation.

\begin{figure*}
    \centering
    \includegraphics[width=\textwidth]{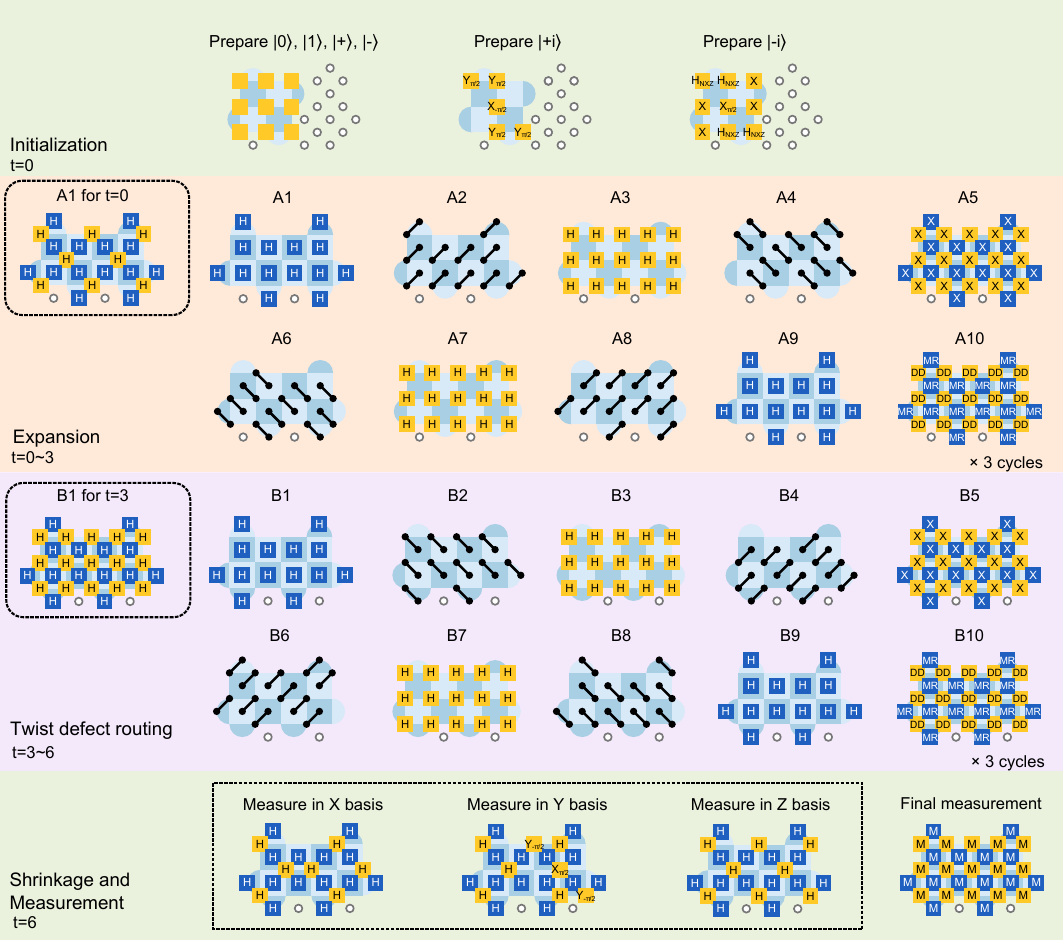}
    \caption{\textbf{Gate schedule for the logical $H$ circuit.}
        The distance-$3$ schedule is shown for the process-tomography configurations used to reconstruct the logical $H$ operation. The shaded bands mark initialization, expansion, twist-defect routing, and shrinkage/readout stages. The top row shows preparation of the $\ket{0}$, $\ket{1}$, $\ket{+}$, $\ket{-}$, $\ket{+i}$ and $\ket{-i}$ logical inputs. Layers A1--A10 and B1--B10 are repeated for three QEC cycles each during expansion and twist-defect routing, respectively; dashed boxes indicate modified boundary layers at stage transitions. Black links denote CZ gates, and site labels denote single-qubit gates, dynamical-decoupling operations, resets or measurements. The final row shows the $X$-, $Y$- and $Z$-basis tomography readout settings compiled with the transversal Hadamard gates on data qubits. The schedule implements the patch rotation with a final Hadamard domain wall, using a $3\times5$ footprint and six syndrome-extraction cycles.}
    \label{fig:CircuitH}
\end{figure*}

The compiled schedule is shown in Fig.~\ref{fig:CircuitH}. The input patch is first expanded into a $d\times(2d-1)$ rectangular footprint and run for $d$ QEC cycles. The logical corners, equivalently the twist-defect world lines, are then routed for another $d$ QEC cycles, implementing the corner permutation $abcd\rightarrow dabc$. Finally, the auxiliary region is shrunk out and a transversal physical-H layer is applied at the output time boundary. The shrink step and the final Hadamard layer do not require an additional syndrome-extraction cycle. Thus the full schedule has depth $2d$ QEC cycles; for the experimental distance-$3$ implementation this gives a $3\times5$ spatial footprint and six syndrome-extraction cycles, matching the operation shown in Fig.~4a of the main text.

Compared with the construction of Ref.~\cite{supp_geherErrorcorrectedHadamardGate2023}, we do not append the final routing step that would return the output patch to its original spatial location. The required logical-movement primitive has already been demonstrated and benchmarked independently in Appendix~G, and can be added when a fixed input--output location is required. Here we omit this extra routing step to isolate the logical-$H$ benchmark from the additional error of a separate movement operation.

The CZ-gate ordering in this dynamic deformation must be chosen more carefully than in a static memory experiment. A single fault on a syndrome ancilla can propagate through the CZ layers into a correlated two-data-qubit hook error. For a static patch, the CZ order can be chosen so that such hooks are perpendicular to the relevant logical operator. During the H-gate deformation, however, the boundary types and the twist-defect locations change in spacetime, so a hook-error direction that is harmless before the deformation can become aligned with a boundary-to-boundary logical error chain after the deformation.

\begin{figure}
    \centering
    \includegraphics[width=0.6\textwidth]{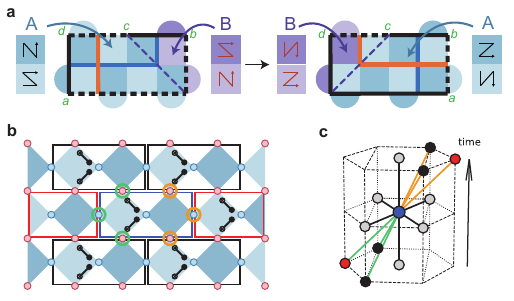}
    \caption{\textbf{Hook-error directionality in the logical $H$ circuit.}
        \textbf{a,} Partition of the rectangular patch into schedule regions A and B during the two halves of the $H$-gate deformation. The logical corners, equivalently twist defects, are labelled $a$--$d$. Moving a twist defect changes the boundary frame, so a hook-error edge could connect a new boundary to a stationary twist defect with the wrong CZ-layer order. \textbf{b,} Mid-cycle detecting regions for a representative stabilizer-measurement cycle. Red and blue circles denote data and ancilla qubits, respectively. Light and dark shaded diamonds show detecting regions whose support expands onto, or shrinks away from, adjacent data qubits during the CZ sequence. Highlighted open circles mark representative faults that connect the central detecting region to neighbouring regions, including distance-two hook errors generated by CZ error propagation. \textbf{c,} Detector-graph representation of the same mechanism. Black and red circles correspond to the detecting regions outlined by the black and red boxes in \textbf{b}, respectively. Green and yellow solid lines correspond to the faults marked by the green and yellow circles in \textbf{b}. The CZ-layer order fixes the temporal direction of hook-error edges; choosing the A and B schedules in Fig.~\ref{fig:CircuitH} prevents these directed edges from concatenating into an undetected boundary-to-twist logical error chain.}
    \label{fig:HookErrorDirection}
\end{figure}

Figure~\ref{fig:HookErrorDirection} summarizes the schedule constraint used to avoid this failure mode. Panel a shows the two CZ-ordering regions used before and after the logical-corner routing. The moving twist defect changes the boundary frame, creating a possible undetectable chain from the new boundary back through the earlier region to a stationary twist defect if the hook-error direction is chosen incorrectly. Panel b illustrates the corresponding mid-cycle detecting regions in a standard surface-code cycle. Faults occurring between CZ layers can connect different detecting regions and form hook errors. Panel c shows the same mechanism in the detector graph: the hook-error edges have a definite temporal propagation direction, set by the CZ-layer order. We choose the A and B schedules in Fig.~\ref{fig:CircuitH} so that these directed hook-error edges cannot concatenate into an undetectable logical chain through the deforming boundary frame.

\subsection{Error analysis}

The decoded tomography data in Fig.~4b,c of the main text verify the intended Pauli flow of the logical $H$ gate. Table~\ref{tab:H_ptm_nonzero} lists the ideal-support PTM elements, together with the calibrated Pauli simulations using the same tomography-boundary SPAM convention as the experiment and the corresponding simulations with tomography SPAM removed. The two non-$Y$ transfer amplitudes, $X\rightarrow Z$ and $Z\rightarrow X$, are $0.7961(18)$ and $0.7841(46)$, respectively, whereas the $Y\rightarrow -Y$ amplitude is lower, $-0.5464(42)$. The off-support PTM elements in the full reconstructed matrix are close to zero, indicating that the dominant reconstructed process is the expected Hadamard Pauli-axis exchange rather than a spurious logical transfer. The tomography-based average logical gate fidelity obtained from the reconstructed PTM is $0.8544(12)$.

\begin{table}[H]
    \centering
    \caption{\textbf{Ideal-support Pauli-transfer-matrix elements of the logical $H$ gate.}
        The table lists the non-zero PTM elements expected for an ideal Hadamard gate, the experimentally reconstructed values, the calibrated Pauli-simulation values including tomography SPAM, and the corresponding simulation values with tomography SPAM removed. Parenthetical uncertainties are 95\% confidence-interval half widths, obtained from five decoding trials for the bootstraped experimental entries or repeated simulation.}
    \label{tab:H_ptm_nonzero}
    \scriptsize
    \begin{ruledtabular}
    \begin{tabular}{cccccc}
        Input & Output & Ideal & Exp. & Sim. & No SPAM \\
        \hline
        $I$ & $I$ & $1.0000$ & $1.0000$ & $1.0000$ & $1.0000$ \\
        $Z$ & $X$ & $1.0000$ & $0.7841(46)$ & $0.8381(30)$ & $0.8637(100)$ \\
        $Y$ & $Y$ & $-1.0000$ & $-0.5464(42)$ & $-0.6175(43)$ & $-0.7765(24)$ \\
        $X$ & $Z$ & $1.0000$ & $0.7961(18)$ & $0.8522(16)$ & $0.8766(25)$ \\
    \end{tabular}
    \end{ruledtabular}
\end{table}

To test whether the reduced $Y\rightarrow -Y$ amplitude is expected from the compiled circuit and the calibrated noise model, we simulated the same $H$-gate tomography circuits using the experimentally calibrated asymmetric Pauli noise model and decoded the results with the NN decoder. As shown in Table~\ref{tab:H_ptm_nonzero}, the simulated ideal-support PTM amplitudes for the two non-$Y$ transfer channels are close to each other, $0.8522(16)$ for $X\rightarrow Z$ and $0.8381(30)$ for $Z\rightarrow X$. In contrast, the simulated $Y\rightarrow -Y$ PTM element is only $-0.6175(43)$. The reconstructed tomography-based average gate fidelity for these SPAM-included simulations is $0.8806(13)$. Thus the same hierarchy observed in the experiment is reproduced by circuit-level simulation: the logical $Y$ channel is expected to be weaker than the $X/Z$ transfer channels even under the calibrated Pauli-noise description.

This $Y$-channel attenuation has two contributions. First, the tomography circuits that access $Y_L$ require non-transversal $Y$-basis state preparation and readout, whereas the $X$- and $Z$-basis tomography settings use transversal surface-code SPAM. Appendix~F shows that $Y$-basis SPAM has lower contrast than transversal $X/Z$-basis SPAM. Second, an encoded $Y_L$ eigenstate is sensitive to logical failures in both the $X_L$ and $Z_L$ channels, so its per-cycle logical decay is larger than that of a single $X$- or $Z$-basis eigenstate. Appendix~F gives an empirical ratio of approximately $1.8$ between the $Y$-basis decay rate and the transversal $X/Z$ decay rates. To separate these two effects, we repeated the $H$-gate Pauli simulations with noiseless boundary QEC cycles at the beginning and end of the circuit, following the convention used to remove tomography-boundary SPAM from the simulated logical channel. In this SPAM-removed simulation, the two $X/Z$ transfer amplitudes increase only slightly, to $0.8766(25)$ and $0.8637(100)$, whereas the magnitude of the $Y\rightarrow -Y$ PTM element increases substantially, to $0.7765(24)$. The corresponding SPAM-removed tomography-based average gate fidelity is $0.9212$. This shows that the lower $Y$-basis SPAM contrast is a major source of the experimental $Y$-channel attenuation. However, even after removing tomography-boundary SPAM, the magnitude of the $Y\rightarrow -Y$ PTM element remains below the $X\rightarrow Z$ and $Z\rightarrow X$ PTM amplitudes. The effective logical error rates extracted from these SPAM-removed PTM amplitudes are $p_X=0.0617$, $p_Y=0.1118$ and $p_Z=0.0682$. The $Y$-channel error rate is therefore larger than the corresponding $X$ and $Z$ rates by factors of $1.81$ and $1.64$, broadly consistent with the independent SPAM-calibration result in Appendix~F.

Together, the experimental PTM, the calibrated Pauli simulations with SPAM, and the SPAM-removed simulations support the interpretation used in the main text: the logical $H$ gate implements the intended Hadamard Pauli flow, while the smaller $Y\rightarrow -Y$ amplitude is dominated by the additional cost of $Y$-basis tomography access and by the greater sensitivity of $Y_L$ to combined $X/Z$ logical failure channels.

\section{Appendix J: Logical S gate}

\subsection{Circuit construction}

We implement the logical $S$ gate using gate teleportation with a fault-tolerantly prepared logical $\ket{+i}$ ancilla. Let the input state be $\ket{\varphi}_D$ on the data patch and the ancilla state be $\ket{+i}_A$. A joint logical $Z_DZ_A$ measurement is followed by an $X_A$ measurement of the ancilla. Under our Pauli-frame convention, the product of these two decoded measurement signs determines whether a virtual $Z_D$ update is applied to the remaining data patch. After this frame update, the output state is $S\ket{\varphi}_D$. Thus, in this gate-teleportation implementation, the key resource is fault-tolerant access to the logical $Y$ basis for preparing the $\ket{+i}$ ancilla.

\begin{figure*}
    \centering
    \includegraphics[width=\textwidth]{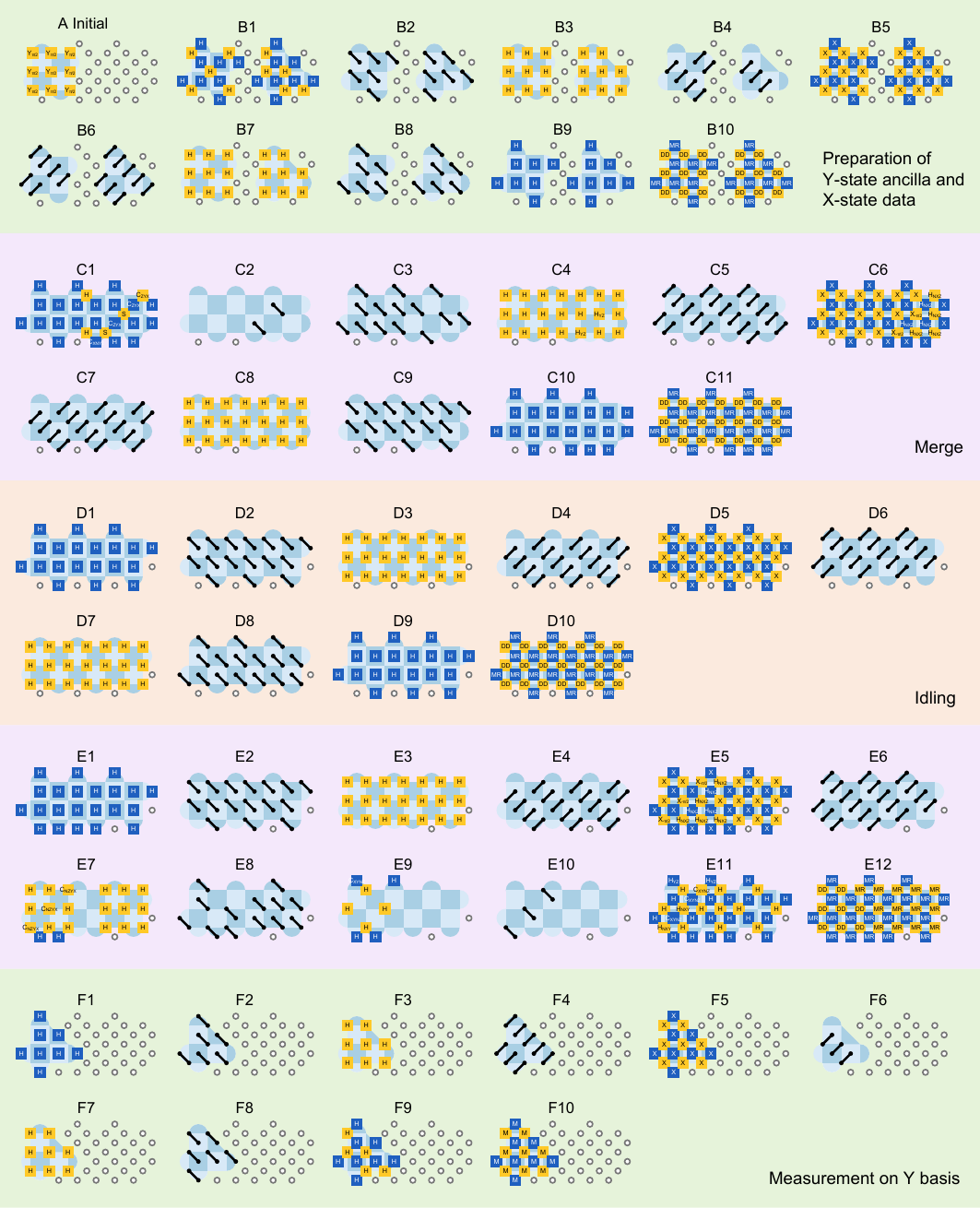}
    \caption{\textbf{Detailed gate schedule for the logical $S$-gate circuit.}
        This distance-$3$ schedule shows a representative tomography configuration with data-patch preparation in the $X$ basis and fault-tolerant $Y$-basis readout. The upper green band prepares the data patch and logical $\ket{+i}$ ancilla, with the ancilla generated by the diagonal-twist $Y$-basis access circuit. The purple and orange bands implement the fused start of the logical $Z_DZ_A$ merge and the subsequent rectangular-patch cycles. For $X$/$Z$ readout, shrinkage and logical-basis measurement are absorbed into the final rectangular-patch cycle; for $Y$ readout, the lower purple and green bands combine shrinkage with fault-tolerant $Y$-basis readout. Black links denote CZ gates; site labels denote single-qubit gates, dynamical-decoupling operations, resets or measurements. The schedule uses a $3\times7$ data-qubit footprint and has depth $4$ QEC cycles for $X$/$Z$ readout or $5$ cycles for $Y$ readout.}
    \label{fig:CircuitS}
\end{figure*}

The compiled distance-$3$ circuit is shown in Fig.~\ref{fig:CircuitS}. The data patch and the $\ket{+i}$ ancilla patch are prepared in parallel. For $X$- and $Z$-basis tomography inputs, the data patch uses transversal surface-code SPAM and waits while the ancilla undergoes fault-tolerant $Y$-basis preparation. For $Y$-basis tomography inputs, the same fault-tolerant $Y$-basis preparation circuit is applied to the data patch and can be run in parallel with the ancilla preparation. The two patches are then merged into a $3\times7$ rectangular patch for the logical $Z_DZ_A$ measurement. The final ancilla readout is implemented by shrinkage rather than by a separate split operation, reducing the spacetime volume of the relevant observable flows.

The $Y$-basis access circuit follows the diagonal-twist construction of Ref.~\cite{supp_gidney2024inplace}. In this construction, a domain-wall surface terminates in the code bulk. The boundary of this surface is the world line of a twist defect, and the resulting diagonal trajectory realizes the twist-defect permutation used for the logical $S$ gate~\cite{supp_brown2017poking,supp_bombin2023logical}. A straightforward implementation of this diagonal-twist access circuit uses one modified QEC cycle for the twist, together with an additional $(d+1)/2$ QEC cycles to make faults generated by the twist detectable.

Our compiled circuit first reduces this count by introducing a horizontal fusion between a hexagonal surface-code region and a standard rectangular region, so that the modified diagonal-twist cycle is merged with an adjacent QEC cycle. The scalable fused schedule therefore has volume $V_S(d)=d\times(2d+1)\times\left[d+\frac{d+1}{2}\right]$. For the experimental distance-$3$ circuit, circuit-distance verification shows that one additional QEC cycle can be removed without reducing the verified distance. The implemented distance-$3$ schedule therefore has spacetime volume $V_S(3)=3\times7\times4$. The verified distance-$3$ implementation occupies a $3\times7$ footprint and uses four QEC cycles for $X$/$Z$ readout, with one additional QEC cycle when the output is read in the fault-tolerant $Y$ basis.

\subsection{Error analysis}

The decoded state-tomography data in Fig.~5c of the main text verify the intended Pauli flow of the logical $S$ gate. The $X$-basis inputs are mapped to the $Y$ basis, with decoded output expectations $\langle Y_L\rangle=0.5844(72)$ for $\ket{+}$ and $-0.5895(44)$ for $\ket{-}$. The $Y$-basis inputs are mapped to the opposite $X$ basis, with $\langle X_L\rangle=-0.5904(62)$ for $\ket{+i}$ and $0.5880(34)$ for $\ket{-i}$. The $Z$-basis inputs are preserved more strongly, with $\langle Z_L\rangle=0.9026(48)$ for $\ket{0}$ and $-0.8986(21)$ for $\ket{1}$.

\begin{table}
    \centering
    \caption{\textbf{Ideal-support Pauli-transfer-matrix elements of the logical $S$ gate.}
        The table lists the non-zero PTM elements expected for an ideal $S$ gate, the experimentally reconstructed values, and calibrated Pauli-simulation values using the same S-gate circuit and tomography-boundary convention as the experiment. Parenthetical uncertainties are 95\% confidence-interval half widths, obtained from decoder fine-tuning trials for the experimental entries and from repeated simulation trials for the simulation entries.}
    \label{tab:S_ptm_nonzero}
    \begin{ruledtabular}
    \begin{tabular}{ccccc}
        Input & Output & Ideal & Exp. & Sim. \\
        \hline
        $I$ & $I$ & $1.0000$ & $1.0000$ & $1.0000$ \\
        $Y$ & $X$ & $-1.0000$ & $-0.5870(30)$ & $-0.6889(24)$ \\
        $X$ & $Y$ & $1.0000$ & $0.5892(29)$ & $0.6942(28)$ \\
        $Z$ & $Z$ & $1.0000$ & $0.9006(29)$ & $0.9309(19)$ \\
    \end{tabular}
    \end{ruledtabular}
\end{table}

The corresponding PTM ideal-support elements are listed in Table~\ref{tab:S_ptm_nonzero}. The dominant entries have the expected signed-permutation structure, $X\rightarrow Y$, $Y\rightarrow -X$ and $Z\rightarrow Z$, and the off-support entries of the reconstructed PTM are close to zero. Using the linear-inversion process matrix and the convention described in Appendix~E, the tomography-based average logical gate fidelity is $0.8461(8)$. The same calibrated Pauli-noise simulation gives an average gate fidelity of $0.8857(7)$, reproducing the hierarchy between the high-contrast $Z\rightarrow Z$ element and the lower $X/Y$-related elements.

This hierarchy follows the observable-flow structure of the gate. The $Z_L$ representative remains on the data patch through the gate-teleportation circuit and does not require a logical $Y$-basis preparation or readout in the tomography boundary. It is therefore protected mainly by the long direction of the intermediate $3\times7$ rectangular patch and by transversal $Z$-basis SPAM. By contrast, the $X_L$ and $Y_L$ transfer channels use the ancilla-mediated part of the teleportation circuit and require fault-tolerant $Y$-basis access either in the input preparation or in the output readout. Appendix~F shows that such $Y$-basis access has a lower fixed contrast and a larger per-cycle decay than transversal $X/Z$-basis SPAM. Unlike in the logical $H$ and CNOT tomography, this $Y$-basis access is not purely a removable tomography-boundary contribution for the $S$ gate, because the fault-tolerant $\ket{+i}$ ancilla is an intrinsic part of the gate itself.

The absolute experimental contrasts are lower than the calibrated Pauli-simulation values, especially for the $X/Y$ transfer channels. This indicates that the simplified independent Pauli model does not capture all error sources in the compiled $S$-gate circuit. One likely contribution is the nonuniform CZ-layer structure of the $S$-gate schedule: some QEC cycles contain five CZ layers, which constrains the placement of the dynamical-decoupling $\pi$ pulses and can reduce the protection against dephasing during those cycles. Other device-specific effects, including leakage and residual crosstalk, may also contribute. We therefore use the Pauli simulation and the simple observable-flow picture as a qualitative consistency check rather than as a complete quantitative error budget. Together, the experimental PTM and simulations support the interpretation used in the main text: the implemented gate realizes the intended logical $S$ Pauli flow, while the reduced $X/Y$ transfer amplitudes arise from the larger ancilla-mediated observable-flow volume and the cost of logical $Y$-basis access.

\section{Appendix K: Circuit-distance verification}
\label{app:circuit_distance_verification}

We define the circuit distance $d_{\rm circ}$ as the minimum number of independent circuit-level fault mechanisms that can cause a logical failure without producing detection events. A compiled logical-operation schedule is distance preserving when $d_{\rm circ}$ reaches the target code distance $d$.

As an exact certificate for the circuits implemented in the experiment, we compiled each distance-$3$ logical-operation circuit to a SAT instance using \texttt{circuit.shortest\_error\_sat\_problem} and minimized the number of selected fault mechanisms with the RC2 MaxSAT solver~\cite{supp_Gidney2021StimAF,supp_krentel1988complexity}.  In all cases, no set of fewer than three independent circuit-level fault mechanisms produced an undetected logical failure. The MaxSAT results therefore verify the experimental distance-$3$ implementations of the logical gates at the full circuit level. 

For the scaled logical CNOT- and $H$-gate schedules, we also checked the graphlike fault distance using Stim~\cite{supp_Gidney2021StimAF}. The detector error models generated from the circuit-level noise model can be decomposed into two hyperedge-free graphlike detector subgraphs, corresponding to the two types of logical observables. In this case, the relevant circuit distance is obtained by finding the minimum-weight graphlike error chain for the logical observables in each subgraph. Applying \texttt{circuit.shortest\_graphlike\_error} to the scaled schedules with odd code distances from $d=3$ to $d=15$ returned minimum graphlike error weights equal to the target distance for every checked CNOT and $H$ circuit. We further used \texttt{circuit.search\_for\_undetectable\_logical\_errors} on the $d=3$ to $d=13$ CNOT and $H$ schedules to confirm that the expected weight-$d$ logical mechanisms set the corresponding upper bound.

The logical $S$-gate schedule requires a separate discussion because its in-place $Y$-basis access implements diagonal twist-defect motion through stabilizer-merging code deformation~\cite{supp_gidney2024inplace}. A corresponding search for undetectable logical errors revealed a timelike $Y$-type hook-error mechanism in the $Y$-basis preparation step. During the stabilizer-merging part of the schedule, indicated in Fig.~\ref{fig:CircuitS}(C2--C3), fault mechanisms affecting two neighboring qubits along the anti-diagonal of the ancilla patch can combine into a local $Y$-type error that triggers one $Z$-type detector and one $X$-type detector. This error can concatenate with a single $ZZ$ fault on an anti-diagonal CZ gate in each preceding idle round, located at the layer shown in Fig.~\ref{fig:CircuitS}(B6), forming a logical error chain that connects the two diagonally moving twist defects.

This hook-error mechanism does not affect the exact distance-$3$ certificate for the experimentally implemented $S$ gate described above. For larger-distance extensions of the fused schedule, however, it indicates that making the $Y$-basis preparation strictly satisfy $d_{\rm circ}=d$ requires $d-2$ idle detection rounds in total before the diagonal fusion of the twist defects in Fig.~\ref{fig:CircuitS}(C). However, related simulations of logical $S$-gate constructions indicate that, when the idle interval reaches $(d+1)/2$ cycles, the effect of such $Y$-type hook-error mechanisms is small and the in-place $Y$-basis schedule still enters a logical-error-suppression regime at low physical error rates~\cite{supp_gidney2024inplace,supp_hiraiSpacetimeVolumeLogicalS2026}. Our distance-scaling simulations in Fig.~\ref{fig:DistanceScaling} show the same behavior for the compiled $S$-gate schedule.

\section{Appendix L: Distance scaling simulation}

To assess how the experimental logical-operation schedules behave at larger code distances and lower physical error rates, we perform circuit-level Pauli simulations for scaled-up versions of the logical $H$, $S$ and CNOT circuits. The simulation uses a symmetric Pauli noise model parameterized by the average physical error rates summarized in Appendix A. We then uniformly rescale all physical error probabilities by a factor $\alpha$, so that $\alpha=1$ corresponds to the present experimental error rates, $\alpha<1$ corresponds to lower physical error rates and $\alpha>1$ corresponds to higher physical error rates. All simulated detection events are decoded using the correlated-matching decoder described in Appendix D.

We use logical process tomography in the simulations to evaluate the same gate-level fidelity metric as in the experiment, while taking advantage of the simulation setting to control the SPAM contribution. For the logical $H$ and CNOT gates, the tomography SPAM operations are treated as ideal boundary operations: one noiseless QEC cycle is inserted at the beginning and end of each circuit, and the $X$/$Y$/$Z$-basis SPAM used for tomography do not contribute stochastic faults. This convention removes the basis-dependent SPAM contribution and isolates the fidelity of the logical gate schedule itself. For the logical $S$ gate, the fault-tolerant preparation of the logical $\ket{+i}$ ancilla is an intrinsic part of the gate-teleportation circuit, so no such boundary modification is applied and the circuit noise is retained throughout the simulated gate schedule.

For each gate and code distance, we simulate the deterministic observables required for logical process tomography using input states selected from $\{\ket{0},\ket{1},\ket{+},\ket{+i}\}$ and measurement bases selected from $\{X,Y,Z\}$. The decoded observables are reconstructed into a physical process matrix by maximum-likelihood estimation, and the plotted logical gate error rate is defined as $p_{\rm L}=1-F_{\rm avg}$, where $F_{\rm avg}$ is the average gate fidelity of the reconstructed logical process. We extract an operation-level error-suppression factor by fitting each distance-scaling curve to $p_{\rm L}(d)=A\Lambda^{-d/2}$. Thus, $\Lambda$ is the fitted reduction factor in logical gate error rate when the code distance is increased by two; $\Lambda>1$ indicates logical-error suppression with increasing distance. For each simulated circuit, samples are accumulated in batches of $10^4$ shots until at least ten logical-error events are observed, which keeps the low-error, large-distance points statistically resolved.

\begin{figure*}
    \centering
    \includegraphics{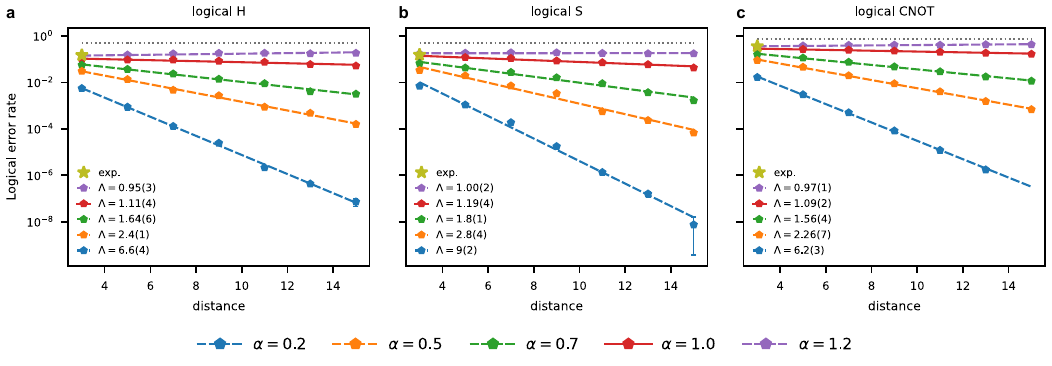}
    \caption{\textbf{Distance-scaling simulations for logical gates.}
        \textbf{a--c,} Simulated logical gate error rate, $p_{\rm L}=1-F_{\rm avg}$, as a function of code distance for the logical $H$ gate (\textbf{a}), $S$ gate (\textbf{b}) and CNOT gate (\textbf{c}). Coloured markers show uniformly rescaled circuit-level Pauli-noise simulations, where the scale factor $\alpha$ multiplies the average physical error probabilities listed in Appendix~A; $\alpha=1$ corresponds to the present calibrated error rates. Yellow stars mark the distance-$3$ experimental process-tomography error rates. Dashed lines are fits to $p_{\rm L}(d)=A\Lambda^{-d/2}$, with fitted error-suppression factors $\Lambda$ shown in the panel legends. The grey dotted line marks the logical gate error rate of a completely depolarizing channel for the corresponding one- or two-qubit Hilbert space. Error bars denote $95\%$ confidence intervals and are smaller than the markers where not visible.}
    \label{fig:DistanceScaling}
\end{figure*}

Figure~\ref{fig:DistanceScaling} shows that, under the symmetric Pauli noise model used here, the present error rates place the operation-level scaling close to the crossover regime, with fitted $\Lambda$ values near unity. Because this model does not include non-Pauli error sources such as leakage and crosstalk, these fitted suppression factors may be higher than would be obtained under full experimental conditions. Uniformly reducing the physical error rates moves the operation-level distance scaling out of this crossover regime and into a clear error-suppression regime: when the physical error probabilities are reduced to half of their present values ($\alpha=0.5$), all three logical gates show clear suppression of the logical gate error rate with increasing code distance. These simulations therefore provide a performance-level consistency check for the scaled gate schedules and support the outlook that improved physical operations would move the implemented logical $H$, $S$ and CNOT circuits into a pronounced logical-error-suppression regime at larger code distances.

\end{document}